\documentclass[journal]{IEEEtran}
\usepackage{amsmath,amssymb,graphicx,booktabs,threeparttable,xcolor,cite,url,amsthm}

\newif\ifrevmarks\revmarksfalse
\DeclareRobustCommand{\rev}[1]{\ifrevmarks\textcolor{red}{#1}\else#1\fi}
\DeclareRobustCommand{\mov}[1]{\ifrevmarks\textcolor{blue}{#1}\else#1\fi}
\newenvironment{revblock}{\ifrevmarks\color{red}\fi}{}
\newenvironment{movblock}{\ifrevmarks\color{blue}\fi}{}
\begin{document}
\title{HOMER: Multi-Person Pose and Box Estimation from Dual Linear-Array Radar in Unseen Rooms}
\author{Haochen~Chai, Qixu~Zhu, and Haobo~Li%
\thanks{The authors are with the School of Science and Engineering, University of Dundee, Dundee DD1 4HN, U.K.}%
\thanks{Corresponding author: Haobo Li.}}
\maketitle
\begin{abstract}
\rev{Millimeter-wave radar can monitor older people at home without a camera, but every home is a room the radar has not seen in training, often shared by several people. We present HOMER, which estimates the pose and bounding box of several people from two linear-array radars and is built on two properties of the measurement that hold in every room. First, the echo of one person extends at most half a body width plus one angular resolution cell from the body center and can hold several peaks; HOMER keeps one peak within this response radius and refines each detection in a search window that shrinks from layer to layer. Second, a camera label covers only the visible part of a person, whereas the radar senses the whole body; HOMER estimates the whole body in metric units, and a separate readout, trained without altering that estimate, moves the bottom edge of the box to where the visible part ends. On the P2S2 split of the MMVR dataset, whose test set includes a room absent from training, HOMER reaches 25.3 box AP in that room against 19.1 for REXO with the same four input frames, a difference of 6.2 with a 95\% interval of [4.3, 7.8]. On the full split it reaches 25.1 against 23.5, and in the unseen room it reaches 16.1 keypoint AP. Code is available at \url{https://github.com/rtb-1005/Homer}.}
\end{abstract}
\begin{IEEEkeywords}
Millimeter-wave radar, ambient assisted living, multi-person pose estimation, person detection, linear array, angular resolution, non-maximum suppression, unseen environments, transformer.
\end{IEEEkeywords}
\section{Introduction}
\label{sec:intro}
\begin{figure*}[t]\centering\includegraphics[width=\textwidth]{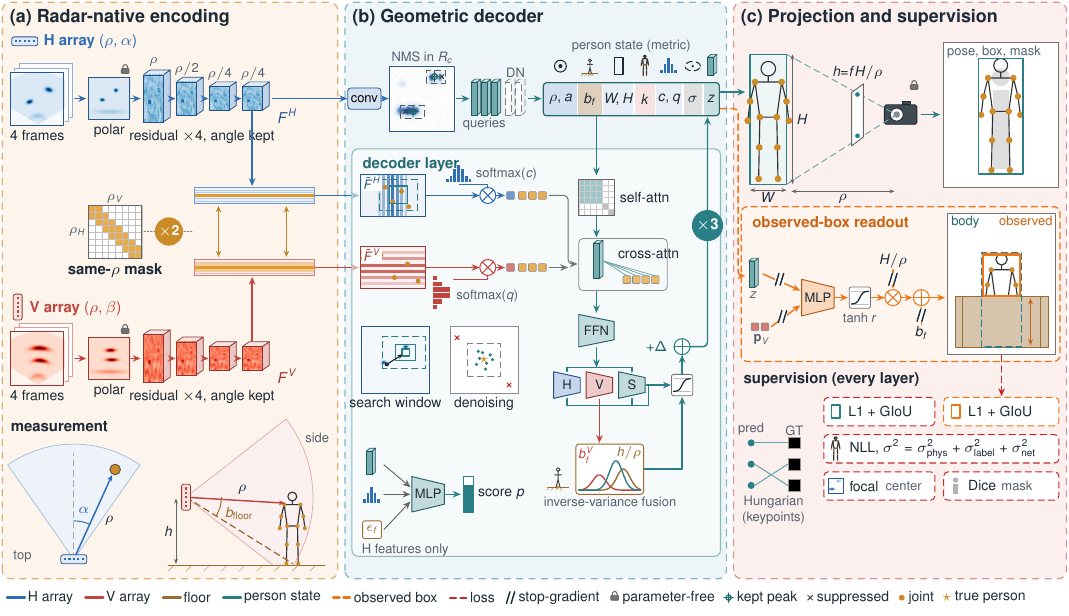}
\caption{Overview of HOMER. (a)~Horizontal (blue) and vertical (red) heatmaps are resampled onto polar grids, encoded, and exchanged only at equal range. (b)~Queries start at peaks kept by suppression at the response radius and are refined in three decoder layers with shrinking windows; each holds a metric body state (teal) whose foot coordinate is fused with the floor prior (brown). (c)~The state is projected into the image as body box, keypoints and mask, and a readout on stop-gradient copies moves only the bottom edge to give the observed box (orange).}\label{fig:framework}\end{figure*}

\begin{revblock}
As the global population ages, one in six people will be 60 or older by 2030~\cite{who_ageing}, making falls one of the most serious health risks of later life. Falls are the second leading cause of unintentional injury death worldwide, and 71\% of the estimated 842{,}000 fall deaths in 2023 occurred among adults aged 60 and over~\cite{who_falls}. Much of the harm comes after the fall itself: most falls among the oldest old happen when they are alone, many cannot get up without help, and lying on the floor for a long time brings pressure sores, dehydration, hypothermia, pneumonia, and a higher risk of death~\cite{fleming2008}. The cost is borne by health services as well as by families, with falls estimated to cost the UK National Health Service more than \pounds 2.3 billion per year~\cite{nice2013falls}.

Continuous, privacy-preserving home monitoring can shorten response times after a fall and help ease this burden. It can also capture slower clinical changes in older adults managing chronic conditions, such as the decline in walking speed strongly linked to cardiovascular mortality~\cite{dumurgier2009}. Both applications demand more than a binary decision over a sensor window: a practical system must track where every person is and how their body is posed---distinguishing a fall from lying down and following gait over weeks---even in a home layout it has never seen.

Existing home-monitoring sensors each trade away a requirement that domestic deployment depends on~\cite{chaccour2017}. Cameras and depth sensors capture full-body pose, yet they place imaging devices in private rooms, generate large streams often processed off-site, and fail in the dark~\cite{cippitelli2017}. Wearable inertial sensors measure motion directly, yet they protect only those who remember to wear and recharge them daily~\cite{ding2025agpose}. WiFi channel state information reuses existing routers, yet its robustness, cross-environment generalization, and privacy remain unresolved~\cite{ma2019wifi}. Instrumented floors, thermal arrays, and passive infrared sensors protect visual privacy, yet they sacrifice spatial detail or require room-by-room installation. Moreover, window-level classifiers built on these modalities often confuse a fall with sitting or lying down quickly~\cite{chaccour2017}.

Millimeter-wave radar avoids these limitations. It forms no image of the person, so there is no picture to leak; it needs nothing on the body and nothing to recharge; its measurements are small enough to be processed, and even learned from, on the device in the home~\cite{wang2026pfedradar}; and it works without light and through clothing and light occlusion~\cite{cippitelli2017}. What radar can report about the people in a room, however, has depended on how its echo is represented. Early systems classified the micro-Doppler signatures of one radar, the time-varying velocities of the limbs, with handcrafted features, and later networks recognized activities and even identified people by their gait from micro-Doppler, range-Doppler and three-dimensional radar cubes, fusing several domains, views or radar nodes~\cite{yang2022gait,dey2024har,yuan2025mfecnet,bao2025tsardc}, in some cases through walls~\cite{zhao2025twhar}. A signature tells what a person does, but not where each person is or how the body is posed, and spatial representations were developed to answer that. Single-chip radars give sparse point clouds that must be accumulated and reconstructed before they describe a body~\cite{yin2026zaxis}, whereas cascaded arrays give dense horizontal and vertical heatmaps, from which RF-Pose~\cite{rfpose} learned keypoints with labels transferred from a synchronized camera. Later work added single- and multi-person pose in two and three dimensions~\cite{hupr,raptr}, including through walls with MIMO radar~\cite{wang2024throughwall,sun2026throughwall}, and query-based detectors that place boxes in three-dimensional radar space~\cite{retr,rexo}.

Most of these advances were made under conditions that a home does not offer. A home is often shared by several people, whose activities must be told apart~\cite{zhao2025lifeactivities}, and every home that installs a radar is a room the sensor has never seen. Activity and gesture recognition meet this shift by domain adaptation from simulated radar data~\cite{hu2025simhar} or by multiview learning that generalizes across domains~\cite{chen2025dggesture}, but both classify what happens rather than locate who is there. MMVR~\cite{mmvr} made multi-person pose and box estimation in an unseen room measurable: it records with two linear arrays and a co-located camera, and its P2S2 split tests on a room absent from training. In that unseen room, the strongest published detectors, RETR~\cite{retr} and REXO~\cite{rexo}, reach only 6.4 and 19.1 box AP. Both leave two physical properties of the sensing setup itself to be learned from data: a linear array's finite angular resolution spreads a single person's echo into several heatmap peaks, on which a peak-based decoder can place several queries, while the co-located camera supervises only the visible part of a person even though the radar senses the whole body.

In this paper, we address both issues directly by building these two room-invariant measurement properties into the architecture, rather than relying on the network to relearn them from room-specific data. We call the resulting framework HOMER (Home-agnostic One-peak Metric-body Estimation from Radar), which estimates the skeleton and bounding box of every person from the two MMVR radars in an unseen room. First, to handle finite angular resolution, a linear array resolves a person laterally only to within one angular cell whose width grows with range, extending the echo one cell beyond the body on either side. We therefore keep at most one peak within the response radius $R_c=W/2+\rho\Delta\alpha$---half a body width plus one angular cell at range $\rho$---and let each query search a window that shrinks from layer to layer. Second, to resolve the visibility mismatch between radar and camera (Section~\ref{sec:signal}), the network maintains a metric state of the whole body from which the box, keypoints, and mask are rendered, while a gradient-detached readout adjusts only the bottom edge of the body box to the visible extent so that truncated camera labels cannot pull the body state toward its visible part (Fig.~\ref{fig:framework}).
\end{revblock}

The main contributions of this paper are as follows.
\begin{enumerate}
\item HOMER, a multi-person framework for dual linear-array radar in which each person is a metric body state, from which the box, keypoints and mask are rendered analytically into the image. In the unseen room of MMVR it reaches $25.32\pm0.73$ box AP from the four input frames of REXO, against 19.08 and 6.40 for the strongest published detectors, REXO and RETR.
\item Peak suppression at a response radius derived from the angular resolution of the array, with search windows that shrink from layer to layer. It cuts duplicate detections in the unseen room from 372 to 70 per 1{,}000 people, and the radius, set without tuning, falls at the knee of the duplicate curve.
\item An observed-box readout detached from the gradient, which lets the body state describe the whole body while the camera label covers only its visible part. It adds 2.53 box AP in the unseen room and 4.46 in the seen sessions, and the gain lies in windows with a person cut by the image border.
\end{enumerate}

\rev{Our code is available at \url{https://github.com/rtb-1005/Homer}. Section~\ref{sec:related} reviews the closest detectors, Section~\ref{sec:method} describes HOMER block by block, Sections~\ref{sec:setup} and~\ref{sec:results} present the experiments, and Section~\ref{sec:conclusion} concludes and outlines future work.}

\section{Related Work}
\label{sec:related}

RETR~\cite{retr} adapts DETR to two radar views, tying them through a shared depth in its positional encoding and a learned radar-to-camera transformation, so that boxes predicted in three dimensions are supervised in the image. REXO~\cite{rexo}, the closest prior work, \rev{lifts diffusion-based box detection} into three-dimensional radar space, uses the noisy boxes to associate features across the two views, and fixes part of the box with the prior that a person touches the ground. The two works differ in how detections are spaced and how the label is read out. REXO lets the denoising process decide where boxes go and refines the projected box as a whole, whereas HOMER keeps at most one peak within the response radius and reads the visible box out of a separate metric body state by moving only its bottom edge. RAPTR~\cite{raptr} addresses three-dimensional pose on HIBER and MMVR under weak supervision, with a two-stage \rev{transformer} decoder; we project its joints into the image for a two-dimensional comparison (Section~\ref{sec:setup}).

\section{Method}
\label{sec:method}
\rev{HOMER follows the three blocks of Fig.~\ref{fig:framework}: radar-native encoding, a geometric decoder, and projection and supervision. Constants come from the released calibration of MMVR~\cite{mmvr}; derivations are in the Supplementary Material.}

\subsection{\rev{Radar-Native Encoding}}
\label{sec:encode}
\label{sec:network}
\rev{In the first block (Fig.~\ref{fig:framework}a), each of the two radars of MMVR~\cite{mmvr} has its linear array along one axis, horizontal ($X$) or vertical ($Y$), and both look along the depth axis $Z$.} \mov{In the radar frame the stored vertical coordinate points down, so the floor lies at a positive height $h$ below the radar. A point $(X,Y,Z)$ is described by the depth-aligned range $\rho=Z$, which is the forward distance rather than the slant range, and by the tangent coordinates
\begin{equation}
a=X/Z,\qquad b=Y/Z,
\label{eq:frames}
\end{equation}
which are the coordinates of the resampled heatmaps and of every person state.}

\begin{movblock}
A linear array resolves the range of a reflector and one direction cosine along its own axis, so the horizontal radar measures $(\rho,\alpha)$ and the vertical radar $(\rho,\beta)$~\cite{vantrees}; range is the only coordinate the two views share. The range cell is $\Delta r=c/2B=11.5$\,cm~\cite{mmvr}. For an $N$-element half-wavelength virtual array the angular cell in direction-cosine space is $2/N$~\cite{vantrees}, and each MMVR radar has 86 virtual elements, 1.3$^\circ$ at boresight~\cite{mmvr}:
\begin{equation}
\Delta\alpha=\Delta\beta=2/86,\qquad \delta x(\rho)=\rho\,\Delta\alpha,
\label{eq:cell}
\end{equation}
where $\delta x$ is the lateral width of one cell at range $\rho$ (Supplementary Sec.~\ref{S-app:resolution}). A range cell is the same everywhere, but a lateral cell grows from 2.3\,cm at 1\,m to 7.0\,cm at 3\,m and 13.4\,cm at 5.75\,m (Fig.~\ref{fig:polar_rc}a). Information should therefore pass between the views only at equal range, and the encoders may coarsen range but not angle.
\end{movblock}

\emph{Encoders.}
Each view contributes a stack of four heatmap frames, the input of REXO~\cite{rexo}, on a logarithmic scale. A fixed, parameter-free operator resamples each stack once onto its own polar grid $(\rho,\text{angle})$ from 0.31 to 5.75\,m, with four samples per angular cell, so no angular resolution is lost at the input (Supplementary Sec.~\ref{S-app:polar}). The two views have separate encoders of identical structure, built from four residual stages with group normalization and 32 to 128 channels; two of the stages stride by two along range only, which reduces range to a feature cell of 12.5\,cm, close to $\Delta r$, and keeps every angle column.

\emph{Exchange at equal range.} The two feature maps are fused by two cross-attention layers in which a row at range $\rho$ in one view attends only to rows within one feature cell of $\rho$ in the other view, each followed by a convolution along range. The attention mask thus encodes the single geometric fact that both arrays measure range. Following REXO~\cite{rexo}, which found the horizontal view to carry most of the detection evidence, the center heatmap, the search score and the person score are computed from horizontal features taken before the exchange, the person score also reading the normalized disagreement $\epsilon_f$ between the vertical foot estimate and the floor prior; the vertical view refines a detected person but does not decide whether one is present.

\subsection{\rev{Geometric Decoder}}
\label{sec:onepeak}
\rev{The second block (Fig.~\ref{fig:framework}b) turns horizontal peaks into people and keeps one query per person.} \mov{People are proposed at peaks of the horizontal center heatmap, trained with the penalty-reduced focal loss of CenterNet~\cite{centernet} on Gaussian targets one cell wide. Each peak gives a query its range $\rho$ and lateral coordinate $a$, and its initial vertical coordinate is the soft-argmax of the vertical heatmap along the row at the same range. Each query carries a person state in metric units: position $(\rho,a,b)$, body size $(\ln W,\ln H)$ initialized to a 0.7\,m by 1.7\,m person, the foot coordinate $b_f$, 17 keypoint fractions $\mathbf{k}=(k^u,k^v)$ inside the body rectangle, 32-bin column and row profiles, a $32\times32$ mask residual, and log-uncertainties for the joints and for range.}

\begin{figure}[t]\centering
\includegraphics[width=0.94\columnwidth]{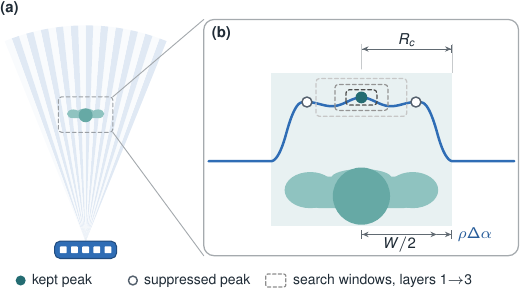}
\caption{Peak suppression at the response radius (schematic). (a)~The angular cells of the horizontal array (alternate shading) widen with range; one person stands in the field of view. (b)~Zoom on that person\rev{, drawn for a uniformly reflecting body}: the echo extends one cell $\rho\Delta\alpha$ beyond the body on either side, \rev{and} its top can hold several peaks. Only the strongest peak within the response radius $R_c=W/2+\rho\Delta\alpha$ (shaded band) is kept, and the kept query searches a window that shrinks from layer to layer. Drawn for $\rho=3$\,m, $W=0.68$\,m and the nominal $\Delta\alpha=2/86$.}\label{fig:polar_rc}\end{figure}

\begin{figure}[t]\centering
\includegraphics[width=\columnwidth]{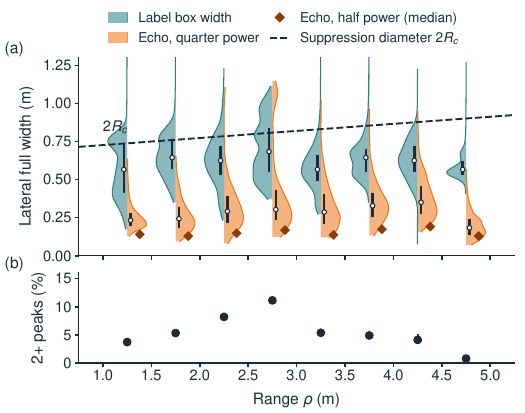}
\caption{\rev{Lateral extent of one person in the raw horizontal heatmap, over 79,787 single-person training and validation frames. (a)~Per 0.5\,m range bin, the label box width (left) and the full width of the echo at a quarter of its peak power (right), with quartiles and median; diamonds: median width at half power; dashed: the suppression diameter $2R_c$. (b)~Share of frames with two or more peaks above half power, with 95\% Wilson intervals.}}\label{fig:extent}\end{figure}

\rev{\emph{Peak suppression at the response radius.}} \rev{A person spans about one body width laterally, and the horizontal array spreads the echo of every part of the body over about one resolution cell on either side~\cite{vantrees}. Measured from the person's center, the main-lobe response therefore lies within the half-extent}
\begin{equation}
R_c=\frac{W}{2}+\rho\,\Delta\alpha ,
\label{eq:rc}
\end{equation}
which we call the response radius (Fig.~\ref{fig:polar_rc}; Supplementary Sec.~\ref{S-app:rc}). With the body width $W=0.68$\,m used by the suppression and search code and $\Delta\alpha=2/86$, it is 0.36\,m at 1\,m, 0.41\,m (about six angular cells) at 3\,m and 0.47\,m at 5.75\,m. \rev{In 79,787 single-person training and validation frames, the label boxes are 0.63\,m wide at the median, close to this $W$, and the raw horizontal heatmap already shows two or more peaks above half power on the one person in 6.9\% of them (Fig.~\ref{fig:extent}).}

\rev{The response of one person} can hold several local maxima of the center heatmap, so taking the top-$K$ local maxima, as a heatmap detector would~\cite{centernet}, often places several proposals inside one person's response. All of them can then be pulled to the same person, and the surplus becomes high-scoring duplicates that a set-based loss~\cite{detr} must learn to suppress from data alone.

We therefore apply peak suppression with the radius \eqref{eq:rc} before the queries are formed. The 64 strongest $3\times3$ local maxima of the horizontal center heatmap are sorted by score and kept greedily: a candidate is discarded when a stronger kept peak lies within $R_c$ laterally, measured in meters as $|\rho_i a_i-\rho_j a_j|$ with $R_c$ evaluated at the candidate's range, and within 0.3\,m in range. The eight highest kept peaks initialize the queries; if fewer than eight survive, the remainder are filled from the discarded candidates in score order. The greedy rule is that of non-maximum suppression; its radius is predicted by the array resolution and the body width rather than tuned. A validation sweep made after the configuration was fixed confirms the prediction: $R_c$ sits at the knee of the duplicate curve; halving it more than triples the duplicates, and radii up to $1.5R_c$ change box AP by at most 0.2 (Supplementary \rev{Fig.~\ref{S-fig:radius}}).

\rev{\emph{Decoder and shrinking search windows.}} \mov{Three decoder layers refine the state. Each layer renders the current box, keypoints and mask, encodes them as a feedback vector, and samples the horizontal features at 32 columns across the person's lateral extent and the vertical features at 32 rows across its vertical extent, pooled with the predicted profiles, as well as at each joint. The query then passes through self-attention among queries, cross-attention to its own sampled tokens and a feed-forward block~\cite{detr}. A horizontal head updates $a$, $\ln W$, $k^u$ and the column profile; a vertical head updates $b_f^{V}$, $\ln H$, $k^v$ and the row profile; a shared head updates $\rho$ and the person score. Increments are bounded per layer to 0.15\,m in range and 0.05 in $a$ and $b$, which keeps each layer's estimate tied to the radar measurement. During training, two groups of denoising queries built from perturbed ground-truth states are added~\cite{dndetr} (Supplementary Sec.~\ref{S-app:training}).}

Suppression at initialization alone does not prevent two queries from converging later. Each decoder layer $l$ therefore moves a query by a soft-argmax over a learned score, sampled from the horizontal features before the exchange, on a window of $\pm1$ range cell and $\pm R_l$ angular cells around its current position, at temperature $0.05\,R_l$. The half-width shrinks from layer to layer,
\begin{equation}
R_1=\min\Bigl(8,\;\frac{R_c}{2\rho\Delta\alpha}\Bigr),\qquad R_2=\sqrt{R_1},\qquad R_3=1,
\label{eq:windows}
\end{equation}
so that at 3\,m the first window reaches $R_c/2$ laterally (2.9 cells), the second 1.7 cells and the last one cell. A soft-argmax is a convex combination of positions inside its window, so two queries at nearly equal range whose kept peaks lie more than $R_c$ apart cannot meet in the first search step. The bounded regression after each search can still bring them together (Supplementary Sec.~\ref{S-app:rc}); the windows make a collapse onto one person less likely without excluding it.

\rev{\emph{Floor prior.}} \mov{A standing person's feet lie on the floor, a horizontal plane at height $h$ below the radar, so in tangent coordinates they lie at
\begin{equation}
b_{\text{floor}}=h/\rho ,
\label{eq:bfloor}
\end{equation}
the foot-contact prior of single-view metrology that REXO also uses~\cite{rexo}. In every decoder layer the vertical branch's foot estimate $b_f^{V}$ is fused with this prior by inverse variance, with $h$ and its uncertainty calibrated from the ankles of training people only and, for the unseen room, set to the median over the training rooms (Supplementary Sec.~\ref{S-app:floor}). Because the uncertainty of the prior falls as $1/\rho^2$ while that of the vertical array is fixed in angle, the prior receives only about 2\% of the weight at 3\,m and 12\% at 5\,m; it is a weak physical prior, and we claim no gain from it.}

\subsection{\rev{Projection and Supervision}}
\label{sec:obsbox}
\rev{The third block (Fig.~\ref{fig:framework}c) maps the metric state into the image, where the labels live.} \mov{A radar-space point $\mathbf{x}$ is mapped to the camera frame with the MMVR extrinsics, $\mathbf{x}_c=\mathbf{R}^{-1}(\mathbf{x}-\mathbf{t})$, and then to pixels by the pinhole model
\begin{equation}
u=c_x+f\,\frac{X_c}{Z_c},\qquad v=c_y+f\,\frac{Y_c}{Z_c},
\label{eq:pinhole}
\end{equation}
with $f=379.476$\,px and $(c_x,c_y)=(322.457,241.627)$ on a $640\times480$ image, followed by one learnable two-dimensional affine transform shared by all frames and initialized from the calibration. Image quantities are therefore computed from a metric person state rather than regressed in pixels. Because the radar and the camera nearly share an origin, one angular cell covers about $f\Delta\alpha\approx8.8$\,px at every distance, so the lateral pixel error is set almost entirely by the angular estimate (Supplementary Sec.~\ref{S-app:projection}).}

\begin{movblock}
The body is a rectangle at depth $\rho$ facing the radar, with corners
\begin{equation}
\Bigl(\rho,\;a\pm\frac{W}{2\rho},\;b_f\Bigr),\qquad \Bigl(\rho,\;a\pm\frac{W}{2\rho},\;b_f-\frac{H}{\rho}\Bigr),
\label{eq:corners}
\end{equation}
so that its bottom edge lies on the foot coordinate. The corners are projected through \eqref{eq:pinhole}, and the body box is the extent of the four projected points. Joint $j$ sits at $\bigl(\rho,\,a-W/(2\rho)+k^u_jW/\rho,\,b_f-k^v_jH/\rho\bigr)$ before projection, and the mask is the outer sum of the column profile, the row profile and the residual. No pixel coordinate is regressed directly. As in REXO~\cite{rexo}, image quantities come from projecting a radar-space state, which every decoder layer corrects from features sampled at the projected geometry; unlike REXO's three-dimensional box, the planar rectangle carries a single range and adds no depth extent to the image box.
\end{movblock}

\begin{figure}[t]\centering
\includegraphics[width=0.94\columnwidth]{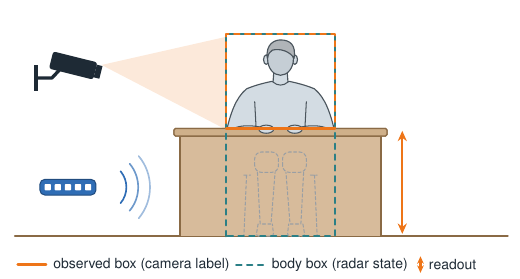}
\caption{The extent mismatch (schematic). For a person seated behind a desk, the camera label covers only the visible upper body, whereas the radar senses the whole body. The body box (teal, dashed) is rendered from the metric body state and reaches the feet; the observed box (orange) shares its top and side edges, and the readout, which reads stop-gradient copies of the state, moves only the bottom edge (arrow).}\label{fig:obs_projection}\end{figure}

\rev{\emph{Observed-box readout.}} Because the label covers only the visible part of a person (Section~\ref{sec:signal}), a single box predicted from the metric state would have to fit both the physical extent that the radar and the floor prior describe and the visible extent that the label describes, and any compromise between them would be carried into the body state and hence into the keypoints. We keep the two apart. The body box is rendered from the metric state as above. A separate readout head predicts the observed box and may change only the bottom edge, which is where the two extents differ when the legs are occluded or cut by the image border (Fig.~\ref{fig:obs_projection}). With $\operatorname{sg}(\cdot)$ denoting the stop-gradient operator, the observed bottom is
\begin{equation}
\begin{aligned}
b_{\text{obs}}&=\operatorname{sg}(b_f)+\operatorname{sg}\!\left(\frac{H}{\rho}\right)\tanh r,\\
r&=\operatorname{MLP}\big(\operatorname{sg}(\mathbf{z}),\operatorname{sg}(\mathbf{p}_V)\big),
\end{aligned}
\label{eq:obsbox}
\end{equation}
where $\mathbf{z}$ is the query feature and $\mathbf{p}_V$ the profile-pooled vertical feature. The bottom edge is projected from $b_{\text{obs}}$ with the same geometry as the body box, and the other three edges are copied, without gradient, from the body box. The MLP is initialized to zero, so the observed box starts equal to the body box, and $|\tanh r|<1$ limits the correction to one body height. Because every input to the head is detached, the observed-box losses reach neither the metric state nor the backbone; of the shared parameters, only the image affine receives a gradient from them (Supplementary Sec.~\ref{S-app:obsbox}), and the observed box is not fed back to later layers.

\rev{\emph{Losses.}} Predictions are matched to people with the Hungarian algorithm on a cost of five times the L1 distance between predicted and labeled visible keypoints minus the person score. Keypoints use a Gaussian negative log-likelihood whose variance is the sum of the image-plane footprint of half an angular cell, the COCO per-joint label variance scaled by the box area~\cite{coco}, and a learned term. The body box uses L1 and generalized IoU, masks a Dice loss on a $32\times32$ crop at the predicted box, the person score binary cross-entropy with weight 0.1 on unmatched queries, and the center heatmaps the focal loss above; all terms are applied to every decoder layer and to the denoising queries. The observed box uses L1, generalized IoU and a smooth-L1 term on its bottom edge in pixels divided by 16. A Gaussian likelihood in radar coordinates on the refined range, angle, width and vertical extent lowers the box AP (Section~\ref{sec:ablation}), and HOMER does not use it. The radar-plane boxes of MMVR still define the targets of the center heatmaps and the states from which the denoising queries are perturbed.

\section{Experimental Setup}
\label{sec:setup}

\subsection{Data and Protocol}
\rev{MMVR~\cite{mmvr} records every scene with the two radars of Section~\ref{sec:encode} and a co-located camera that supplies the labels (Supplementary Fig.~\ref{S-fig:frames}).}

\begin{figure}[t]\centering\includegraphics[width=0.94\columnwidth]{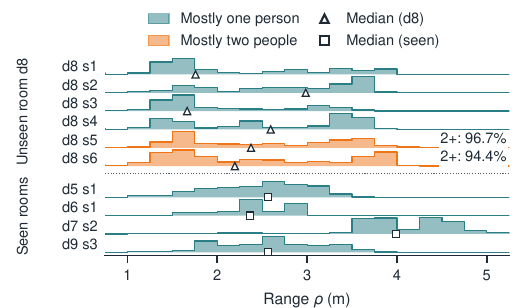}
\caption{\rev{Range of the people in the ten P2S2 test sessions: the six sessions of the unseen room d8 (52,115 windows) and the four held-out sessions of the seen rooms (33,562 windows). Histograms of $\rho$ over all labeled people (0.25\,m bins) with their medians; $\rho$ is the range-axis center of the radar box, a proxy for the slant range. Orange: sessions where most windows hold two or more people, with that share.}}\label{fig:data}\end{figure}

We use the P2 part of MMVR~\cite{mmvr}, recorded in several rooms with one to three people. Recordings are labeled $d X s Y$, where $d$ is the recording day, one room per day, and $s$ the session within that day; each session is cut into segments. We follow the official P2S2 protocol. Training uses 135 segments from sessions of rooms d5, d6, d7 and d9, and validation uses 39 segments from four further sessions of the same rooms (d5s2, d6s4, d7s5, d9s6). The test split has 99 segments and 85,677 windows in two parts. The first is room d8, which appears in neither training nor validation: 60 segments and 52,115 windows over six sessions, of which s1--s4 are single-person and s5--s6 contain two people in about 95\% of their windows. The second consists of one whole session from each training room (d5s1, d6s1, d7s2, d9s3), held out from training and validation, with 39 segments and 33,562 windows; it contains a single person in at least 99.9\% of its windows, whereas three of the four validation sessions of the same rooms mostly contain two or three people (Fig.~\ref{fig:data}; Supplementary Table~\ref{S-tab:data}). No session is shared between training, validation and test. We report the two parts separately, as \emph{unseen room} and \emph{seen rooms, held-out sessions}, together with the full split, which is the scope of the published P2S2 numbers of MMVR and REXO. For model selection we use 3,003 windows drawn evenly from the 39 validation segments.

\begin{figure*}[t]\centering\includegraphics[width=\textwidth]{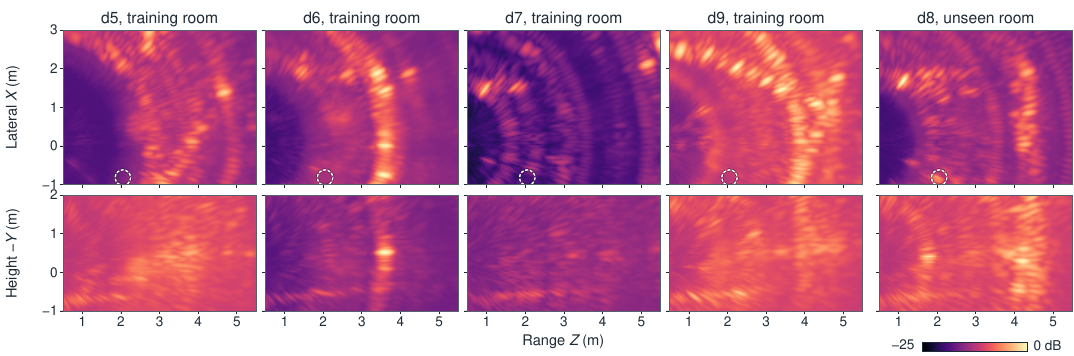}
\caption{\rev{The five rooms as the radar sees them: per-pixel median of the horizontal (top) and vertical (bottom) heatmaps over 10,000--12,000 frames of each room, sampled from all its sessions without reading any label, in dB below the maximum of each map; people are not removed. The dashed circle marks the same place, $X=-0.82$\,m and $Z=2.05$\,m, in every room: a reflector that is bright only in the unseen room d8 (Supplementary Sec.~\ref{S-app:traces}).}}\label{fig:rooms}\end{figure*}

\rev{\emph{Rooms.} The five rooms of P2 differ in size, furniture and radar position, and each leaves its own static pattern in every frame. Fig.~\ref{fig:rooms} shows this pattern as the per-pixel median of the heatmaps of each room, taken without reading any label. The medians differ from room to room in both views, and d8 holds a strong reflector where the training rooms hold none, at $X=-0.82$\,m and $Z=2.05$\,m: it is among the strongest 1\% of the median in all six d8 sessions, 7.7--9.1\,dB above its median, and lies between the 20th and 55th percentiles in the four training rooms. Section~\ref{sec:duplicates} returns to it.}

\rev{\emph{Labels.}}
\label{sec:signal}
\mov{The box that MMVR~\cite{mmvr} provides is the bounding box of the person's mask in the image. It covers what the camera sees, not the physical body: when the legs are hidden behind furniture or leave the frame, the label ends at the visible part.} \rev{On the validation sessions, 19,801 of the 57,743 labeled person instances that do not touch the image border (34\%) have a visible ankle more than 5\% of the box height below the box bottom, so their label box stops short of the feet. The share depends on the room, from 1\% in d6 to 43\% in d9 (Supplementary Fig.~\ref{S-fig:labels}).} \mov{This is the extent mismatch addressed in Section~\ref{sec:obsbox}.}

\subsection{Metrics}
Boxes are scored with COCO average precision~\cite{coco} at IoU thresholds from 0.50 to 0.95 (AP), and at 0.50 and 0.75 (AP$_{50}$, AP$_{75}$). Our primary box output is the observed box clipped to the $640\times480$ image, because the label box is itself confined to the image; detector baselines are clipped by the same rule. Pose is scored with keypoint AP in two forms. The first follows MMVR Eq.~(3)~\cite{mmvr}: predictions are matched to people per frame by the mean object keypoint similarity (OKS) over visible joints, and AP is computed per joint and averaged over the 17 COCO joints. Since that equation does not fix the scale inside OKS, we report it with the label box area and with the label mask area. The second, COCO-12, is the standard COCO keypoint AP over the twelve body joints with the mask area, greedy matching by person score and per-person OKS. As the official MMVR pose evaluator has not been released, both are our implementations.

\subsection{Baselines}
We compare boxes with RETR~\cite{retr} and REXO~\cite{rexo}, using the released weights and scoring their outputs with our evaluator and the same clipping rule. Our evaluator reproduces REXO's published full-split AP (23.51 against 23.47~\cite[Table~1]{rexo}); REXO's diffusion sampler is stochastic, so we average three inference seeds, whose standard deviation is at most 0.09 AP. Clipping leaves REXO's scores unchanged, since its boxes already lie inside the image. REXO and RETR predict boxes only, and no other method has published pose results on this split under a stated OKS scale. RAPTR~\cite{raptr} predicts three-dimensional joints. We project those of its released weights into the image through our calibration, with the trained affine transform of one of our models, and score them with our pose evaluator. Subtracting a constant image offset fixed before the test raises its full-split pose AP from 0.34 to 0.89.

\subsection{Statistics and Test Discipline}
Each configuration is trained with three seeds, and we report the mean and the sample standard deviation over seeds. For the unseen-room difference to REXO we use a paired bootstrap that resamples the 60 d8 segments as contiguous blocks~\cite{kunsch} 2,000 times and recomputes global AP for both methods on each resample.

Earlier development versions of the model were evaluated on d8 for diagnostic purposes, so the unseen room was not unseen by the authors during development. Epochs, thresholds and post-processing were all chosen on validation, and each configuration, the epoch-selection rule and the evaluation code were fixed and recorded before its models were scored. Predictions were frozen and hashed before the test labels were read, and each reported model was scored on the test split once. An audit of the inference code shows that, besides constants fixed in training (the weights, the calibration and the floor constants), the prediction for a window reads only the radar heatmaps of its input frames, and that no label file is opened. Replaying inference on 5,000 fixed windows of the unseen room from these inputs reproduced all eight stored output fields of two of the twelve-frame models bit for bit.

\section{\rev{Results and Discussion}}
\label{sec:results}

\begin{table*}[t]
\centering
\caption{Pose and Box Results on the MMVR P2S2 Test Split. All Rows Are Scored With the Same Evaluator.}
\label{tab:main}
\begin{threeparttable}
\begin{tabular}{@{}l ccc ccc c c@{}}
\toprule
 & \multicolumn{3}{c}{Pose AP (MMVR Eq.~(3), box-area OKS)} & \multicolumn{3}{c}{Box, unseen room} & Box & Box \\
\cmidrule(lr){2-4}\cmidrule(lr){5-7}
Method & Unseen & Seen & All & AP & AP$_{50}$ & AP$_{75}$ & Seen AP & All AP \\
\midrule
RETR~\cite{retr}\tnote{a} & -- & -- & -- & 6.40 & 25.59 & 2.00 & 22.30 & 12.20 \\
REXO~\cite{rexo}\tnote{a,b} & -- & -- & -- & 19.08 & 56.84 & 6.65 & \textbf{31.75} & 23.51 \\
RAPTR~\cite{raptr}\tnote{c} & 0.30 & 0.36 & 0.34 & -- & -- & -- & -- & -- \\
\midrule
HOMER (3 seeds) & \textbf{16.14}$\pm$0.88 & \textbf{17.84}$\pm$1.05 & \textbf{16.52}$\pm$0.87 & \textbf{25.32}$\pm$0.73 & \textbf{68.51}$\pm$2.08 & \textbf{12.23}$\pm$1.40 & 24.44$\pm$0.94 & \textbf{25.10}$\pm$0.69 \\
\quad 12 input frames & 15.68$\pm$0.36 & 17.75$\pm$0.77 & 16.18$\pm$0.49 & 26.58$\pm$1.49 & 68.60$\pm$1.58 & 15.41$\pm$2.78 & 24.20$\pm$0.38 & 25.73$\pm$0.88 \\
\bottomrule
\end{tabular}
\begin{tablenotes}\footnotesize
\item Unseen: room d8 (52,115 windows). Seen: four held-out sessions of the training rooms (33,562 windows). All: the official 85,677-window split. Box AP uses the observed box clipped to $640\times480$; detector boxes are clipped by the same rule. HOMER uses the four input frames of REXO; the last row, an ablation (Table~\ref{tab:ablation}), uses 12 and is otherwise unchanged. Bold: best among the compared methods. Mean $\pm$ sample s.d.\ over training seeds; --: not applicable.
\item[a] Released weights scored with our evaluator and clipping rule. REXO's published full-split AP is 23.47. RETR's published full-split AP is 12.45 in the first arXiv version (the value REXO cites) and 12.19 in later versions.
\item[b] Mean of three diffusion inferences (s.d.\ at most 0.09).
\item[c] Released weights; three-dimensional joints projected into the image through our calibration and scored with our evaluator, which is not RAPTR's original evaluation; its boxes are not scored.
\end{tablenotes}
\end{threeparttable}
\end{table*}

\subsection{Boxes}
\label{sec:boxes}
Table~\ref{tab:main} gives the box results in the three scopes. In the unseen room HOMER reaches $25.32\pm0.73$ AP against 19.08 for REXO, which reads the same four frames, and 6.40 for RETR. The paired bootstrap places the difference to REXO, $+6.24$ AP, in the 95\% interval $[+4.32,+7.80]$. Each of the three seeds, at 25.45, 24.54 and 25.98, is above every REXO inference (19.06--19.10). The relative gain is largest at the strict threshold, where AP$_{75}$ is 12.23 against 6.65. On the full official split HOMER reaches $25.10\pm0.69$ against 23.51, a difference of $+1.58$ $[-0.92,+3.85]$, with all three seeds above the best REXO inference at 23.61. On the held-out sessions of the seen rooms the order reverses, at $24.44\pm0.94$ against 31.75; these sessions are single-person and come from rooms that both models saw in training (Section~\rev{\ref{sec:duplicates}}). \mov{HOMER has 3.5\,M parameters against 17.7\,M for REXO and a counted forward cost of 81.6\,GFLOPs against 5.5.}

\mov{The lead in the unseen room is not carried by a few sessions or by crowding. HOMER is ahead of REXO in all six sessions of the unseen room, by $+0.66$ to $+9.26$ AP, with intervals above zero in three (\rev{Fig.~\ref{fig:strata}}). The lead holds in single-person windows of that room, $+3.73$ $[+1.51,+5.70]$, and grows with two or more people, $+8.82$ $[+6.65,+10.84]$, whereas single-person windows of the seen sessions give $-7.30$ $[-11.73,-2.77]$, so crowding alone does not explain the difference between the two parts. Range separates them further: beyond 4\,m REXO is far ahead, at 48.72 against 27.92 AP for the twelve-frame variant, and such ranges are rare in the unseen room, where the 90th percentile of range is below 4\,m in every session, but make up about half of the seen session d7s2 (\rev{Figs.~\ref{fig:data} and~\ref{fig:strata};} Supplementary Table~\ref{S-tab:data}). On the validation sessions the model is ahead in multi-person windows, $+3.54$ $[+0.50,+7.03]$, and level in single-person windows.}

\subsection{Pose}
\rev{Fig.~\ref{fig:readout}b} reports pose AP under the three conventions of Section~\ref{sec:setup}. With the same OKS area the two matching rules agree closely in the unseen room (11.16 and 11.43) and on the full split (10.79 and 10.30), so the differences between conventions come mainly from the choice of area, the box area giving the more lenient scale~\cite{coco}. \mov{The pose numbers depend on the OKS scale, which MMVR leaves unspecified~\cite{mmvr}; we therefore report three conventions and do not rank against the published RF-Pose number~\cite{rfpose}.}

\begin{figure*}[t]\centering\includegraphics[width=\textwidth]{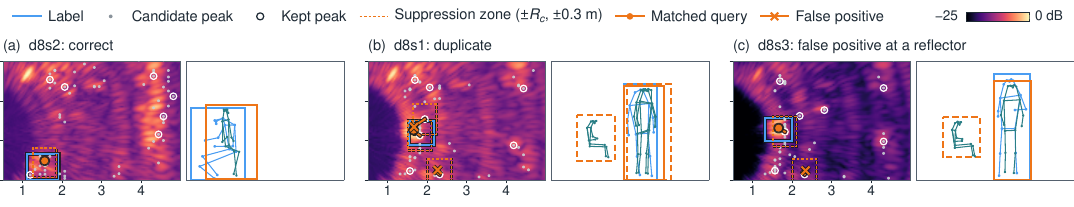}
\caption{Peak suppression on three of the twelve windows of the unseen room fixed before testing, first seed of the twelve-frame variant of Table~\ref{tab:ablation}. Left of each pair: horizontal heatmap\rev{ in dB below the frame maximum}, range (ticks at 1--4\,m) against lateral position, with the 64 candidate peaks, the eight kept peaks and, for each query scoring above 0.5, its suppression zone and its path through the three decoder layers. Right: image plane with the labels (blue) and our boxes and joints (orange, teal); false positives are dashed. (a)~One person keeps one peak, and its query converges on the person. (b)~Two peaks kept 0.41\,m apart, just beyond $R_c=0.38$\,m, are merged by the decoder. (b) and (c)~A false positive at the same unlabeled reflector in two sessions.}\label{fig:suppression}\end{figure*}

\subsection{Ablation}
\label{sec:ablation}
\begin{table}[t]
\centering
\caption{Ablation: Input Length and Radar-Space Supervision (3 Seeds), and Components Removed One at a Time (Seed 1)}
\label{tab:ablation}
\setlength{\tabcolsep}{3pt}
\begin{tabular}{@{}l cc cc c@{}}
\toprule
 & \multicolumn{2}{c}{Validation} & \multicolumn{2}{c}{Unseen room} & All \\
\cmidrule(lr){2-3}\cmidrule(lr){4-5}
Configuration & Pose & Box & Pose & Box & Box \\
\midrule
\multicolumn{6}{@{}l}{\emph{Mean of three seeds}}\\
HOMER (4 frames) & 12.91 & 21.37 & 16.14 & 25.32 & 25.10 \\
\ 12 input frames & 13.10 & 21.58 & 15.68 & 26.58 & 25.73 \\
\ 12 frames, radar-space superv. & 12.43 & 18.25 & 16.10 & 23.76 & 22.40 \\
\midrule
\multicolumn{6}{@{}l}{\emph{Seed 1, 12 frames, radar-space supervision}}\\
All components & 12.67 & 19.49 & 17.75 & 25.41 & 23.69 \\
\ w/o horizontal-only detection & 12.86 & 18.50 & 17.82 & 24.62 & 22.65 \\
\ w/o peak suppression & 6.34 & 13.83 & 7.78 & 20.92 & 19.67 \\
\ w/o floor fusion & 12.16 & 19.02 & 15.26 & 24.47 & 23.16 \\
\bottomrule
\end{tabular}

\vspace{2pt}{\footnotesize Pose: MMVR Eq.~(3), box area. Box: observed box, clipped. Seed s.d.\ on validation: 0.76 (pose) and 1.21 (box) for HOMER, 1.21 and 1.11 with 12 frames and radar-space supervision.}
\end{table}

Table~\ref{tab:ablation} varies the input and the supervision with three seeds each and removes single components with one seed. Twelve input frames instead of four change validation pose and box AP by $+0.19$ and $+0.21$ and unseen-room box AP by $+1.26$, less than the seed standard deviation of the twelve-frame model in that room (1.49), while unseen-room pose AP falls by 0.46; eight more frames buy no reliable gain, and HOMER reads the four frames of REXO. Adding the Gaussian likelihood in radar coordinates of Section~\ref{sec:method} to the twelve-frame model lowers box AP, paired by seed, by 3.33 on validation and by 2.82 in the unseen room, while pose AP changes by $-0.67$ and $+0.42$. The single-component removals start from this twelve-frame, radar-supervised configuration. Removing the peak suppression of Section~\ref{sec:onepeak} has by far the largest effect: with everything else unchanged, unseen-room pose AP falls from 17.75 to 7.78 and box AP from 25.41 to 20.92, and validation pose AP falls from 12.67 to 6.34, well below the weakest seed of that configuration. Restricting detection to the horizontal view and fusing the floor prior change the validation scores by at most 1.0 point, within the seed spread, and we keep both as physical priors. Every row of the table stays above REXO's 19.08 box AP in the unseen room.

\subsection{Contribution of the Observed-Box Readout}
\label{sec:readout}
\begin{figure}[t]\centering\includegraphics[width=\columnwidth]{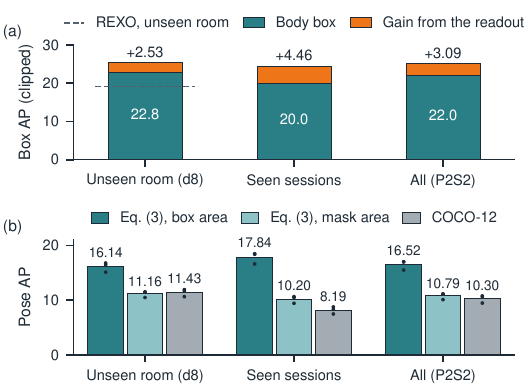}
\caption{\rev{HOMER in the unseen room, the seen sessions and the full split, mean of three seeds.} (a)~Clipped body-box AP (teal) and, stacked on it, the gain from the readout, i.e.\ observed-box AP minus body-box AP (orange); the dashed line marks REXO in the unseen room. \rev{(b)~Pose AP under the three conventions of Section~\ref{sec:setup}; dots: the three seeds.}}\label{fig:readout}\end{figure}

\begin{figure}[t]\centering\includegraphics[width=\columnwidth]{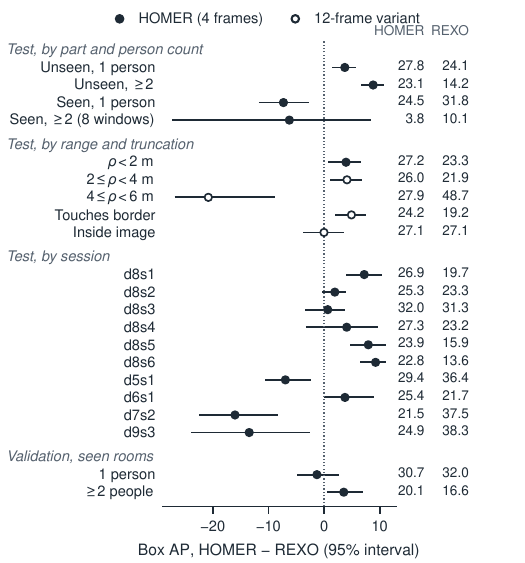}
\caption{\rev{Box AP of HOMER minus REXO by stratum of the test split and of the validation sessions of the seen rooms, with 95\% paired block-bootstrap intervals; the AP of both at right. Hollow: the twelve-frame variant, for strata scored only for it (Supplementary Sec.~\ref{S-app:strata}).}}\label{fig:strata}\end{figure}

Because the readout changes only the bottom edge, comparing the observed box with the body box isolates its contribution (Fig.~\ref{fig:readout}). In the unseen room the clipped body box reaches 22.79 AP and the observed box 25.32, so the readout adds 2.53 AP, and the body box alone is still 3.7 AP above REXO; the readout adds 4.46 in the seen sessions and 3.09 on the full split. With radar-space supervision the body box is nearly unchanged, at 22.99 with twelve frames against 23.12 without it, but the readout adds only 0.77 instead of 3.45, so nearly all of the box AP that this supervision costs in Table~\ref{tab:ablation} is lost in the readout. Split by truncation over both test parts, the lead of the twelve-frame variant over REXO lies in windows where a person touches the image border, $+4.92$ $[+1.99,+7.55]$, while windows with everyone inside the image are level, $-0.02$ (\rev{Fig.~\ref{fig:strata}}).

\subsection{Duplicate Detections}
\label{sec:duplicates}
The suppression is meant to leave one confident detection per person, and Fig.~\ref{fig:suppression} shows it doing so on the radar heatmap: of 64 candidate peaks, one survives on each person, and the query that starts on it converges on that person. Supplementary Table~\ref{S-tab:counts} counts the boxes scoring above 0.5 and calls a duplicate any such box beyond the first that overlaps the same labeled person at IoU above 0.5. Against the same seed without suppression, the pair of Table~\ref{tab:ablation}, in the unseen room duplicates fall from 372 to 70 per 1,000 people and false positives from 73,574 to 46,262, the share of windows with as many confident boxes as people rises from 30.4\% to 52.7\%, and the number of people matched at IoU 0.5 hardly changes (48,862 without and 49,918 with it). The lead of HOMER in the unseen room comes from finding more people, 52,164 of 69,026 against 46,638 for REXO, and from tighter boxes, while REXO keeps fewer confident boxes per person, 1.09 against 1.45. \rev{Where REXO is ahead, in the single-person sessions of the training rooms, HOMER more often adds a second box on the one person, from two peaks on the flanks of one response or from a response beyond $R_c$ (Fig.~\ref{fig:suppression}b and c; Supplementary Table~\ref{S-tab:counts}). A response beyond $R_c$ can also come from the room itself, as in the d8 windows of Fig.~\ref{fig:suppression}b and c, where the false positive sits on the reflector that stands out in the d8 medians of Fig.~\ref{fig:rooms}; a model trained in other rooms has had no chance to learn to ignore it.}

\section{\rev{Conclusion and Future Work}}
\label{sec:conclusion}
Two facts of the measurement, that one person's echo extends half a body width plus one angular cell from its center and that the camera labels only what it can see, led to two simple designs: keep one peak within that radius, and read the visible box out of a whole-body state without letting the label disturb the body. With them, HOMER reaches 25.3 box AP against 19.1 for REXO in the one room of MMVR that is absent from training, from the same four input frames and with a fifth of REXO's parameters.

\begin{revblock}
Knowing where each person is and how the body is posed, in a room the model has never seen, is where recognizing activity and detecting falls at home begins, and our future work builds on it in four directions. First, we will recognize everyday activities such as drinking, sitting down and falling from sequences of the estimated whole-body states, bringing the temporal and multidomain models developed for radar activity recognition~\cite{dey2024har,yuan2025mfecnet,bao2025tsardc} onto a representation that already separates people, so that a fall can be told apart from lying down. Second, we will follow the posture and gait of patients with chronic conditions over weeks, since slow walking is strongly associated with cardiovascular death in older adults~\cite{dumurgier2009}, and combine them with breathing and heartbeat measured by the same radar~\cite{jang2024vital}. Third, we will test in more rooms and real homes, including through walls and in larger spaces~\cite{wang2024throughwall,sun2026throughwall}, and combine the measurement priors of HOMER with domain generalization~\cite{hu2025simhar,chen2025dggesture}. Fourth, we will run the model on the device in the home and keep learning there~\cite{wang2026pfedradar}, so that no radar data have to leave it.
\end{revblock}

\section*{Acknowledgment}
The authors used the following AI systems in preparing this work. Grok (xAI) drafted the text of the manuscript, the figures and the derivations of the Supplementary Material from the authors' code, results and instructions. Claude (Anthropic) reviewed the experimental results and revised the text of the manuscript. Codex (OpenAI) executed the training, evaluation and data synchronization runs on the authors' servers. H.~Chai designed the model, wrote its code and ran the experiments; Q.~Zhu organized the literature and typeset the manuscript; H.~Li supervised the project and guided its direction. All numbers were checked against the logged results, and the authors take full responsibility for the content.
\bibliographystyle{IEEEtran}
\bibliography{refs}

\clearpage
\setcounter{section}{0}\renewcommand{\thesection}{S-\Roman{section}}
\setcounter{figure}{0}\renewcommand{\thefigure}{S\arabic{figure}}
\setcounter{table}{0}\renewcommand{\thetable}{S\arabic{table}}
\setcounter{equation}{0}\renewcommand{\theequation}{S\arabic{equation}}
\renewcommand{\topfraction}{0.9}\renewcommand{\bottomfraction}{0.8}\renewcommand{\textfraction}{0.07}
\renewcommand{\floatpagefraction}{0.85}\renewcommand{\dbltopfraction}{0.9}\renewcommand{\dblfloatpagefraction}{0.85}
\setcounter{topnumber}{3}\setcounter{bottomnumber}{2}\setcounter{totalnumber}{5}
\twocolumn[{\centering{\LARGE Supplementary Material}\par\vspace{14pt}}]
This supplement collects the derivations behind \rev{Section~\ref{sec:method}} of the main paper and the experimental details and additional results that the main paper leaves out. Equations, figures and tables of the main paper are cited with their main-paper numbers; those of the supplement carry the prefix S.
\makeatletter
\@ifundefined{proposition}{\newtheorem{proposition}{Proposition}}{}
\makeatother

\section{Sensor Setup and Frames}
\label{S-app:frames}
\begin{figure*}[t]\centering
\includegraphics[width=\textwidth]{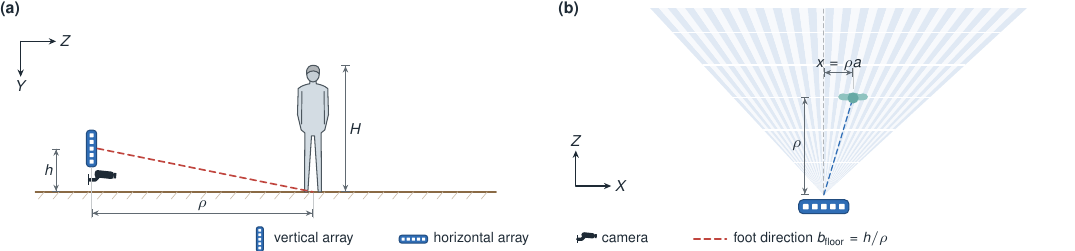}
\caption{Sensor setup and frames. (a)~Side view: the vertical array at the radar origin, the camera offset by the MMVR translation $\mathbf{t}$, the floor at height $h$ below the radar (0.545--0.607\,m per room in the stored frame, whose vertical axis points down), and a person of height $H=1.7$\,m at range $\rho$ whose feet lie in the direction $b_{\text{floor}}=h/\rho$. (b)~Top view: the horizontal array and the depth-aligned grid, with rays of constant $a=X/Z$ (alternate shading, every second angular cell) and lines of constant $\rho=Z$ (every meter) up to 5.75\,m; a person at $\rho=3$\,m and $a=0.3$ lies at $x=\rho a$.}\label{S-fig:frames}\end{figure*}

Fig.~\ref{S-fig:frames} shows the sensors and the frames of \rev{Section~\ref{sec:encode}} inside the room.

\section{Angular and Range Resolution of a Linear Array}
\label{S-app:resolution}
Consider a uniform linear array of $N$ elements at half-wavelength spacing along the $X$ axis. A far-field reflector in direction $\theta$ from boresight produces the element phases $\pi n u$, $n=0,\dots,N-1$, with the direction cosine $u=\sin\theta$~\cite{vantrees}. Steering the array to $u_0$ and summing gives the array factor
\begin{equation}
|A(u)|=\left|\sum_{n=0}^{N-1}e^{j\pi n(u-u_0)}\right|=\left|\frac{\sin\bigl(N\pi(u-u_0)/2\bigr)}{\sin\bigl(\pi(u-u_0)/2\bigr)}\right| .
\label{S-eq:af}
\end{equation}
The numerator vanishes first at $|u-u_0|=2/N$, so the main lobe extends one distance $2/N$ to either side of its peak. By the Rayleigh criterion two equal reflectors are resolved when the peak of one falls on the first null of the other, which gives the cell
\begin{equation}
\Delta u=2/N .
\label{S-eq:rayleigh}
\end{equation}
The half-power width of \eqref{S-eq:af} is about $0.886\cdot2/N$, so \eqref{S-eq:rayleigh} is the conservative, null-to-peak definition. For a MIMO radar the same argument holds for the virtual array formed by all transmit--receive pairs, whose aperture sets $N$. In MMVR each view is recorded by a cascade of four single-chip radars whose virtual array has 86 half-wavelength-spaced elements along the array axis, with an angular resolution of 1.3$^\circ$ at boresight~\cite[Sec.~3.2]{mmvr}. With $N=86$, \eqref{S-eq:rayleigh} gives $\Delta u=2/86=0.0233$, or 1.33$^\circ$, the value $\Delta\alpha=\Delta\beta$ used throughout. Two parts of the cell remain uncertain. MMVR does not state whether an amplitude taper was applied before the angle transform; a taper widens the main lobe~\cite{harris1978}, so $\Delta\alpha$ may understate the true cell. And the cell is constant only in direction cosine, whereas our grid is uniform in tangent, so off axis the effective cell is wider than $\Delta\alpha$, as quantified below.

Range is resolved by the frequency sweep. For bandwidth $B$ the range cell is $\Delta r=c/2B$; MMVR reports $\Delta r=11.5$\,cm~\cite{mmvr}, which corresponds to $B\approx1.30$\,GHz. We do not use the Doppler dimension.

Our grid uses the tangent $a=X/Z$ rather than the direction cosine. The two are related by $u=a/\sqrt{1+a^2}$, so $\mathrm{d}u/\mathrm{d}a=(1+a^2)^{-3/2}$ and one cell in $u$ spans $\Delta a=(1+a^2)^{3/2}\Delta u$ in tangent. We use $\Delta\alpha$ directly as the tangent cell. This is exact at boresight and underestimates the cell off axis, by a factor of 1.40 at $a=0.5$; all radii derived from $\Delta\alpha$ are therefore slightly too small away from the center of the field of view.

\section{From the Polar Grid to Metric Space}
\label{S-app:polar}
With the depth-aligned convention of (\ref{eq:frames}), a grid point $(\rho,a,b)$ maps to
\begin{equation}
X=\rho a,\qquad Y=\rho b,\qquad Z=\rho .
\label{S-eq:metric}
\end{equation}
For the horizontal view the Jacobian of $(\rho,a)\mapsto(X,Z)$ is $\bigl[\begin{smallmatrix}a&\rho\\1&0\end{smallmatrix}\bigr]$ with determinant $-\rho$. A cell of size $\Delta r\times\Delta\alpha$ therefore has the metric area $\rho\,\Delta r\,\Delta\alpha$; its lateral side is $\rho\Delta\alpha$ and its range side is $\Delta r$. With the nominal constants the lateral side is 2.3, 7.0 and 13.4\,cm at 1, 3 and 5.75\,m, against a constant 11.5\,cm in range, so the cell is narrower than it is deep below about 4.9\,m and wider beyond. The vertical view is identical with $(Y,b,\Delta\beta)$ in place of $(X,a,\Delta\alpha)$.

The resampled input grid has 175 range samples from 0.3125 to 5.75\,m at 3.125\,cm, and 348 tangent samples from $-0.930$ to $1.087$ at $\Delta\alpha/4$. Range is oversampled by a factor of 3.7 and angle by 4. The encoders reduce range by four, to a feature cell of 12.5\,cm, close to $\Delta r$, and keep every tangent column, so the feature map still holds four samples per angular cell. Features are read at continuous $(\rho,a)$ by bilinear interpolation on this grid.

\section{Camera Model and Analytic Projection}
\label{S-app:projection}
The physical radar frame has its vertical axis pointing up, while the stored coordinate $b$ points down, so a state $(\rho,a,b)$ is the physical point $\mathbf{x}=(\rho a,\,-\rho b,\,\rho)$. With the MMVR extrinsics $(\mathbf{R},\mathbf{t})$ and intrinsics $(f,c_x,c_y)$, the projection is
\begin{equation}
\begin{gathered}
\mathbf{x}_c=\mathbf{R}^{-1}(\mathbf{x}-\mathbf{t}),\qquad
\begin{pmatrix}u\\v\end{pmatrix}=\mathbf{A}\begin{pmatrix}u'\\v'\\1\end{pmatrix},\\
\begin{pmatrix}u'\\v'\end{pmatrix}=\frac{f}{Z_c}\begin{pmatrix}X_c\\Y_c\end{pmatrix}+\begin{pmatrix}c_x\\c_y\end{pmatrix},
\end{gathered}
\label{S-eq:fullproj}
\end{equation}
where $f=379.476$\,px for both axes, $(c_x,c_y)=(322.457,241.627)$, $\mathbf{t}=(-0.052,0.358,0.096)$\,m, and $\mathbf{A}$ is a $2\times3$ affine transform initialized from the MMVR calibration and learned jointly with the network. No sensor was recalibrated. At initialization the rows of the linear part of $\mathbf{A}$ have norms $s_u=1.050$ and $s_v=0.950$. After training, no entry of the linear part has moved by more than 0.066 in any of the three seeds of the 12-frame variant (Table~\ref{tab:ablation}) and the row norms are 0.98--1.00 and 0.93, so at 3\,m a body of fixed metric size projects 4.9--6.3\% narrower and about 2\% shorter than with the initial transform, and the image of the principal ray moves by 2.0--5.3\,px horizontally and 6.5--7.9\,px vertically. We bound $Z_c$ below by 0.1\,m, which matters only for invalid initial states.

\emph{First-order behavior.} Write $\mathbf{R}=\mathbf{F}\mathbf{Q}$ with $\mathbf{F}=\operatorname{diag}(1,-1,1)$. The residual rotation $\mathbf{Q}$ is 4.8$^\circ$ from the identity. Setting $\mathbf{Q}=\mathbf{I}$ and $\mathbf{A}$ to the identity gives $X_c=\rho a-t_x$ and $Z_c=\rho-t_z$, hence
\begin{equation}
\frac{\partial u'}{\partial a}=\frac{f\rho}{\rho-t_z},\qquad
\frac{\partial u'}{\partial \rho}=\frac{f\,(a\,t_z-t_x)}{(\rho-t_z)^2},
\label{S-eq:jac}
\end{equation}
and the analogous expressions for $v'$ with $(b,t_y)$. The derivative with respect to the tangent is $f$ up to the factor $\rho/(\rho-t_z)$, which is 1.10 at 1\,m and 1.03 at 3\,m. The derivative with respect to range is suppressed by $|\mathbf{t}|/\rho$ relative to it. Evaluated numerically through \eqref{S-eq:fullproj} with the full $\mathbf{R}$ and the initial $\mathbf{A}$, $\partial u/\partial a$ evaluated at $a=b=0$ is 430, 410 and 406\,px at 1, 3 and 5\,m (423, 403 and 399\,px at $b=0.2$). At $a=0$, $b=0.2$, $\partial v/\partial\rho$ is 154, 15.5 and 5.5\,px/m at the same ranges, dominated by the vertical offset $t_y=0.358$\,m between radar and camera.

Two consequences follow. A lateral metric error $\delta X$ at range $\rho$ maps to about $f s_u\,\delta X/\rho$ pixels. One angular cell, $\delta X=\rho\Delta\alpha$, therefore maps to $f s_u\Delta\alpha$ pixels at every range: 9.3\,px with the initial affine, 8.7--8.8\,px with the trained ones and 8.8\,px without an affine. A range error of one cell, $\delta\rho=\Delta r$, moves a point by $\Delta r\,\partial v/\partial\rho$, about 1.8\,px at 3\,m and 18\,px at 1\,m. Close to the sensor, range errors are therefore not negligible in the vertical image direction; this is one reason the encoders keep a range cell close to $\Delta r$ instead of striding further.

\emph{Box and joints.} The four corners (\ref{eq:corners}) are projected by \eqref{S-eq:fullproj}, and the body box that is supervised and evaluated in HOMER is the extent of the four projected points. The simpler size rule of the code, used when the floor module is off and to initialize the log-size of denoising queries, takes the radar range as depth, $w=f s_u W/\rho$ and $h=f s_v H/\rho$. For the initial body size of 0.7 by 1.7\,m at 3\,m this gives 93 by 204\,px with the initial affine and 87--88 by 200\,px with the trained ones. Using the camera depth $Z_c\approx\rho-t_z$ instead would enlarge both sides by the factor $\rho/(\rho-t_z)$, that is by 3.3\% at 3\,m, 5.0\% at 2\,m and 10.6\% at 1\,m, so the range-as-depth rule underestimates the size by that much. The corner extent also includes the rotation and the shear of $\mathbf{A}$, initialized as $\mathbf{A}_0=\left[\begin{smallmatrix}1.0494&-0.0466&-33.82\\0.0307&0.9495&41.74\end{smallmatrix}\right]$. With $\mathbf{A}_0$ the shear of the first row dominates, $-0.047$ times a box height of about 210\,px or roughly 10\,px, and the extent is about 108 by 214\,px at 3\,m. Training turns this shear to between $+0.012$ and $+0.019$ in those seeds, so that it adds at most 4\,px, and keeps both offsets within 0.09\,px of their initial values. Because the rendered extent depends on $\mathbf{A}$, training changes the box through the learned affine as well as through the state. Unlike the projection of the eight corners of a three-dimensional box, which adds the depth extent and enlarges the image box~\cite{rexo}, the planar rectangle carries a single range. For sampling, the projected box edges and joints are mapped back to tangent coordinates at the query's range by inverting \eqref{S-eq:fullproj} along the camera ray; the round-trip error of this inverse is below $2\times10^{-4}$\,px on the calibration audit.

\section{Floor Prior, Calibration and Fusion}
\label{S-app:floor}
\emph{Model.} We assume that the floor is a plane at stored height $h$ below the radar and parallel to its $X$--$Z$ plane, and that the lowest visible ankle of a standing person lies on it. A tilt of the floor relative to the radar is not modeled and is absorbed in the residual spread below. By \eqref{S-eq:metric} a foot at range $\rho$ then has the tangent coordinate $b_{\text{floor}}=h/\rho$ of (\ref{eq:bfloor}).

\emph{Calibration.} The constants are estimated per training room without any plane fit and without recalibrating the camera:
\begin{enumerate}
\item For every associated person in the training segments of the room, take the lowest visible ankle in the label.
\item Back-project that ankle through the fixed MMVR camera geometry, using the inverse of \eqref{S-eq:fullproj}, to the person's labeled range, and read its stored vertical coordinate.
\item Take the median over all such people; this median is $h$.
\item Compute $\sigma_h$ as the standard deviation of 1,000 bootstrap medians that resample whole recording segments~\cite{kunsch}.
\end{enumerate}
The third constant $\sigma_s$ converts the scatter of the ankle rows into an equivalent range error. Since $v\approx c_y+f h_c/Z$ for a camera at height $h_c$ above the floor, a pixel residual $\delta v$ about the fitted row corresponds to the depth error $\delta Z=\delta v\,Z^2/(f\,|h_c|)$, and $\sigma_s$ is the standard deviation of this quantity. It includes pose, label and model error, which is why it is large. Table~\ref{S-tab:floor} lists the constants: $h$ lies between 0.545 and 0.607\,m and $\sigma_s$ between 1.13 and 2.31\,m. Each seen room uses its own constants, and for the unseen room d8, which has no training data, we use the equal-weight medians over the four training rooms, $(h,\sigma_h,\sigma_s)=(0.5879,0.0110,1.3639)$. No validation or test label enters any constant.

\begin{table}[!tbp]\centering
\caption{Per-room floor constants from the training segments.}\label{S-tab:floor}
\footnotesize\setlength{\tabcolsep}{3pt}
\begin{tabular}{lccccc}
\toprule
Room & $h$ (m) & 95\% interval (m) & $\sigma_h$ (m) & $\sigma_s$ (m) & Segments\\
\midrule
d5 & 0.607 & 0.588--0.626 & 0.0099 & 1.13 & 35\\
d6 & 0.591 & 0.573--0.614 & 0.0106 & 1.58 & 31\\
d7 & 0.545 & 0.521--0.565 & 0.0114 & 2.31 & 30\\
d9 & 0.585 & 0.552--0.614 & 0.0165 & 1.15 & 39\\
\midrule
d8 (median) & 0.588 & -- & 0.0110 & 1.36 & --\\
\bottomrule
\end{tabular}
\end{table}

\emph{Uncertainty of the prior.} Differentiating $b_{\text{floor}}=h/\rho$ gives $\delta b=-h\,\delta\rho/\rho^2+\delta h/\rho$. With independent errors of variance $\sigma_\rho^2=(\Delta r/2)^2+\sigma_s^2$ in range and $\sigma_h^2$ in height,
\begin{equation}
\sigma^2_{\text{floor}}=\Bigl(\frac{h}{\rho^2}\Bigr)^2\Bigl(\frac{\Delta r^2}{4}+\sigma_s^2\Bigr)+\frac{\sigma_h^2}{\rho^2}.
\label{S-eq:sfloor}
\end{equation}

\emph{Fusion.} In every decoder layer the vertical branch's estimate $b_f^V$ is fused with the prior. Treating the two as independent Gaussian measurements of the same foot coordinate, the product of the two densities is Gaussian with the inverse-variance mean
\begin{equation}
b_f=\frac{\sigma^2_{\text{floor}}\,b_f^{V}+\sigma^2_{V}\,b_{\text{floor}}}{\sigma^2_{\text{floor}}+\sigma^2_{V}}
\label{S-eq:fusion}
\end{equation}
and variance $(\sigma_V^{-2}+\sigma_{\text{floor}}^{-2})^{-1}$, where $\sigma_V^2$ combines half a resolution cell with a learned term. The weight on the prior is
\begin{equation}
w_{\text{floor}}=\frac{\sigma_V^2}{\sigma_V^2+\sigma^2_{\text{floor}}},\qquad \sigma_V^2=\Bigl(\frac{\Delta\beta}{2}\Bigr)^2+\sigma^2_{\text{net}} .
\label{S-eq:wfloor}
\end{equation}
Because $\sigma_{\text{floor}}$ falls as $1/\rho^2$ while $\sigma_V$ is constant in tangent, the weight rises with range. With the median constants and $\sigma_{\text{net}}=0$, $w_{\text{floor}}$ is 0.3\%, 1.7\%, 5.1\%, 11.6\% and 18.6\% at 2, 3, 4, 5 and 5.75\,m. A learned $\sigma_{\text{net}}>0$ raises every value.

\section{Response Radius and Search Windows}
\label{S-app:rc}
\emph{Lateral response of a person.} Model a person at lateral position $x_0$ as a reflector of uniform strength over $|x-x_0|\le W/2$, and the array point-spread at range $\rho$ by the main lobe of \eqref{S-eq:af}, which in metric units extends $\rho\Delta\alpha$ to either side of its peak. Sidelobes, multipath and the nonuniform reflectivity of the body are ignored. The response is the convolution of the two, and the support of a convolution is the sum of the supports, so the main-lobe response of the person occupies
\begin{equation}
|x-x_0|<\frac{W}{2}+\rho\Delta\alpha=R_c .
\label{S-eq:rcapp}
\end{equation}
Inside $|x-x_0|\le W/2-\rho\Delta\alpha$ the response is close to flat, so several local maxima of a heatmap trained on it can appear there; outside $R_c$ the person contributes nothing through the main lobe. We use the nominal $W=0.68$\,m and $\Delta\alpha=2/86$. The suppression and the search windows use $W=0.68$\,m, whereas the decoder's initial box width is 0.7\,m; the 2\,cm difference changes $R_c$ by 1\,cm.

\rev{\emph{Measured extent.} Fig.~\ref{fig:extent} of the main paper measures the lateral extent of one person in the raw horizontal heatmaps of the 79,787 single-person frames of the training and validation sessions. For each frame the lateral profile is the strongest linear power over the depth rows of the label box, and the echo widths are full widths at half and at a quarter of the peak above the median of the profile. The label boxes are 0.63\,m wide at the median, close to the nominal $W$. The echo is narrower than the uniform model, as expected from a body that does not reflect uniformly: its half-power width has a median of 0.15\,m and its quarter-power width 0.27\,m. The suppression relies on the support of the response rather than its shape, and 96\% of the quarter-power widths and 80\% of the label widths lie within the diameter $2R_c$. Two or more peaks above half power, separated by a valley of at least 1\,dB, occur in 6.9\% of the frames, most often (11\%) at 2.5--3\,m.}

\emph{Suppression.} Candidates are sorted by score, and candidate $j$ is discarded when a kept candidate $i$ satisfies $|\rho_ia_i-\rho_ja_j|<0.34+\rho_j\Delta\alpha$ and $|\rho_i-\rho_j|<0.3$\,m. A kept peak at the person's center removes every other maximum of that person, since the whole main-lobe response lies within $R_c$ of the center. A kept peak on a flank does not: the flat top is $W-2\rho\Delta\alpha$ wide, wider than $R_c$ at every range below $W/(6\Delta\alpha)\approx4.9$\,m (0.60 against 0.38\,m at 1.7\,m), so a second maximum on the opposite flank can survive. Panel (b) of Fig.~\ref{fig:suppression} in the main paper shows one such case. Two different people whose centers are within $R_c$ laterally and within 0.3\,m in range are merged as well, which is the price of the rule.

\emph{Windows.} In layer $l$ the search evaluates the score on the offsets $\{-1,0,1\}\Delta r$ in range and $k\Delta\alpha$, $|k|\le R_l$, in tangent, with $R_l$ from (\ref{eq:windows}), and returns the softmax-weighted mean of these positions.

\emph{Separation and its limits.} Consider two kept queries entering the first decoder layer at $(\rho_i,a_i)$ and $(\rho_j,a_j)$ with $|\rho_i-\rho_j|<0.3$\,m, so that the suppression rule applied to them and guarantees a lateral separation $\rho|a_i-a_j|\ge R_c$. Each search output is a convex combination of window positions within $R_1\Delta\alpha$ of the query's tangent, and $R_1\Delta\alpha\le R_c/(2\rho)$ by (\ref{eq:windows}), so after the first search step
\begin{equation}
\rho|a_i'-a_j'|\;\ge\;\rho|a_i-a_j|-R_c\;\ge\;0 ,
\label{S-eq:sepbound}
\end{equation}
which is a triangle inequality; the factor $\tfrac12$ in $R_1$ is what keeps it non-negative. The bound covers the search step only. The regression increment, at most 0.05 in tangent per layer and query, is not limited by the windows and lets two queries close by about 0.9\,m over three layers at 3\,m, so a complete layer carries no separation guarantee; queries filled from discarded candidates and kept peaks 0.3\,m or more apart in range are not covered either, and the equal-range premise is approximate because each radius is evaluated at its own query's range. Separation also does not place two queries on different people: two queries more than $R_c$ apart can both lie inside one wide response, one on each flank. The bound inherits the nominal $\Delta\alpha$, its use as the tangent cell (Sec.~\ref{S-app:resolution}), and a response model without taper, sidelobes or multipath; a taper widens the main lobe~\cite{harris1978} and would make $R_c$ too small. Finally, $R_c$ describes where a query starts, not how far it can be corrected: a query whose center lies within $R_c$ of a person starts on that person's main-lobe response, while a query farther away still sees the person at the edges of its 32 sampling columns, which span $a\pm W/(2\rho)$, and of its first window. How large an error the trained decoder can correct has not been measured, and the windows have not been varied.

\begin{figure}[!tbp]\centering
\includegraphics[width=\columnwidth]{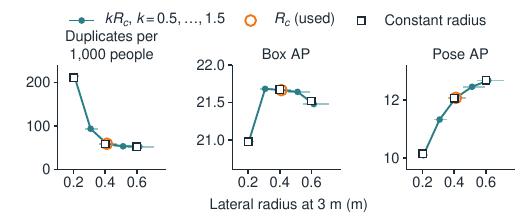}
\caption{\rev{Suppression radius on validation, seed 1 of the 12-frame variant (3,003 windows, 5,950 labeled people). Only the lateral radius of the suppression changes; the search windows, the 0.3\,m range gate and the weights are those of the trained model. Radii proportional to $R_c$ are drawn at their value at 3\,m, with a bar over their values from 1 to 5.75\,m; constant radii (squares) at their value. Box AP: clipped observed box; pose AP: MMVR Eq.~(3) with box area; duplicates as in Table~\ref{S-tab:counts}.}}\label{S-fig:radius}\end{figure}

\emph{Radius sweep.} After the configuration and the test predictions had been fixed, we varied only the lateral radius of the suppression on the validation windows, for seed 1 of the 12-frame variant (\rev{Fig.~\ref{S-fig:radius}}). Below $R_c$ the duplicates rise steeply: at $0.5R_c$ there are 215 per 1,000 people against 59, and box and pose AP fall by 0.7 and 2.0. Above $R_c$ the duplicates fall only to 53, box AP changes by at most 0.2, and pose AP rises by up to 0.6 at $1.5R_c$. Constant radii give the same picture, with 0.4\,m indistinguishable from $R_c$ and 0.2\,m as poor as $0.5R_c$. The sweep places $R_c$ where the duplicates stop falling steeply; it does not show that a larger radius would be worse on these windows, nor that the range dependence of $R_c$ matters at the validation ranges. The test configuration is unchanged.

\section{Geometry and Gradients of the Observed-Box Readout}
\label{S-app:obsbox}
\begin{figure}[!tbp]\centering
\includegraphics[width=\columnwidth]{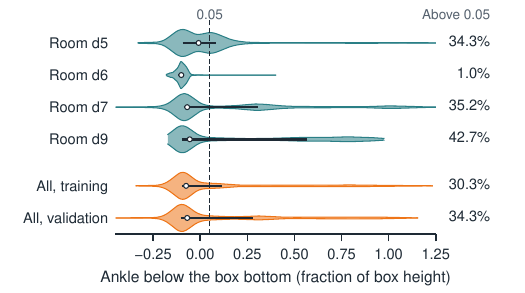}
\caption{\rev{Lower visible ankle below the bottom edge of the label box, as a fraction of the box height, for labeled people inside the image with an associated pose (training and validation sessions; test labels were not used). Violins with quartiles and median, cut at the 0.5th and 99.5th percentiles. Right: share of all boxes inside the image whose ankle lies more than 0.05 below the box bottom.}}\label{S-fig:labels}\end{figure}

\rev{\emph{How far the labels stop short.} Fig.~\ref{S-fig:labels} gives, for every labeled person of the training and validation sessions whose box does not touch the image border and whose pose could be associated with the box, how far the lower visible ankle lies below the bottom edge of the box. Most label boxes end just below the ankles, at the feet, but a long tail ends well above them, and the share beyond 5\% of the box height, the statistic of Section~\ref{sec:signal} of the main paper, ranges from 1\% in d6 to 43\% in d9. In images with several people the pose and box arrays of MMVR are not always stored in the same order, so each box is paired with a pose by the Hungarian matching used in training, between the box and the bounding box of the pose's visible joints (at least three joints, IoU of at least 0.3); the 2,437 validation boxes inside the image left without a reliable pose count as not stopping short, and counting them all the other way would raise the validation share from 34.3\% to 38.5\%.}

\emph{Geometry.} Let $v(b)$ denote the larger image row of the two bottom corners $(\rho,a\pm W/(2\rho),b)$ projected by \eqref{S-eq:fullproj}. The body box has bottom row $y_2=v(b_f)$. The readout predicts
\begin{equation}
\begin{aligned}
b_{\text{obs}}&=\operatorname{sg}(b_f)+\operatorname{sg}(H/\rho)\tanh r,\\
y_2^{\text{obs}}&=\operatorname{sg}(y_2)+v(b_{\text{obs}})-v(b_f),
\end{aligned}
\label{S-eq:obsapp}
\end{equation}
and copies $x_1$, $y_1$ and $x_2$ from the body box without gradient. Since $|\tanh r|<1$, the observed bottom stays strictly between the body top $b_f-H/\rho$ and one body height below the feet. To first order $v(b_{\text{obs}})-v(b_f)\approx(\partial v/\partial b)(H/\rho)\tanh r$, with $\partial v/\partial b\approx356$\,px at 3\,m. The difference form makes the observed box exactly equal to the body box when $r=0$, which is how the zero-initialized head starts.

\emph{Gradients.} Let $\mathcal{L}$ be any loss on the observed box. The inputs to the readout MLP are detached, and $\rho$, $a$, $W$, $H$ and $b_f$ enter \eqref{S-eq:obsapp} only through detached copies, so $\partial\mathcal{L}/\partial\boldsymbol{\theta}=0$ for every parameter of the backbone, the exchange, the decoder layers and the state heads, and the body box, keypoints and mask receive no gradient from $\mathcal{L}$ through the state. Two parameter sets do receive gradient: the readout MLP, through $\partial\mathcal{L}/\partial r=\mathcal{L}'\,(\partial v/\partial b)(H/\rho)\operatorname{sech}^2 r$, and the linear part of the learnable affine $\mathbf{A}$, through which $v(\cdot)$ is evaluated. The affine offset cancels in the difference, and the gradient on the second row of the linear part is proportional to the difference of the projected bottom points at $b_{\text{obs}}$ and $b_f$, so it vanishes at $r=0$ and is first order in $\tanh r$. Since $\mathbf{A}$ also maps the body box and the joints, this is the only path by which the observed-box loss can reach the keypoints. In the 12-frame seeds, training moves each entry of the linear part of $\mathbf{A}$ by at most 0.066 (Sec.~\ref{S-app:projection}), which bounds the combined effect of all losses on it; we have not separated the share of the observed-box loss. The decoder's feedback vector is built from the body box, so the observed box does not enter later layers.

\section{Data Composition and Training Details}
\label{S-app:training}
Table~\ref{S-tab:data} gives the composition of every P2S2 test and validation session used in Section~\ref{sec:setup} of the main paper.

\begin{table*}[t]
\centering
\caption{Composition of the P2S2 Test and Validation Sessions}
\label{S-tab:data}
\setlength{\tabcolsep}{2.5pt}
\footnotesize
\begin{tabular}{@{}ll r ccc cc c@{}}
\toprule
 & & & \multicolumn{3}{c}{Windows with $n$ people (\%)} & \multicolumn{2}{c}{People (\%)} & $\rho$ (m) \\
\cmidrule(lr){4-6}\cmidrule(lr){7-8}
Part & Session & Windows & 1 & 2 & 3 & Border & Legs hidden & p10 / p50 / p90 \\
\midrule
Unseen & d8s1 & 8,623 & 99.99 & 0.01 & 0 & 56.9 & 4.2 & 1.29 / 1.76 / 3.71 \\
Unseen & d8s2 & 8,177 & 100 & 0 & 0 & 26.1 & 16.1 & 1.52 / 2.98 / 3.63 \\
Unseen & d8s3 & 8,758 & 100 & 0 & 0 & 64.5 & 4.1 & 1.34 / 1.66 / 3.22 \\
Unseen & d8s4 & 8,872 & 99.97 & 0.03 & 0 & 32.3 & 8.6 & 1.38 / 2.59 / 3.57 \\
Unseen & d8s5 & 8,880 & 3.28 & 96.70 & 0.02 & 41.9 & 34.8 & 1.48 / 2.38 / 3.65 \\
Unseen & d8s6 & 8,805 & 5.60 & 94.36 & 0.05 & 47.9 & 29.3 & 1.32 / 2.20 / 3.83 \\
\midrule
Seen & d5s1 & 7,808 & 99.94 & 0.05 & 0.01 & 10.7 & 24.9 & 1.71 / 2.56 / 3.16 \\
Seen & d6s1 & 8,592 & 100 & 0 & 0 & 14.0 & 1.4 & 1.75 / 2.37 / 2.92 \\
Seen & d7s2 & 8,392 & 99.96 & 0.04 & 0 & 2.1 & 22.8 & 3.53 / 3.99 / 4.70 \\
Seen & d9s3 & 8,770 & 100 & 0 & 0 & 21.4 & 24.6 & 1.78 / 2.56 / 3.29 \\
\midrule
Val. & d5s2 & 693 & 100 & 0 & 0 & 7.4 & 34.3 & 1.71 / 2.66 / 3.16 \\
Val. & d6s4 & 770 & 6.88 & 92.60 & 0.26 & 44.3 & 15.5 & 0.80 / 2.00 / 2.89 \\
Val. & d7s5 & 770 & 0.26 & 99.48 & 0.26 & 1.5 & 60.6 & 2.98 / 3.70 / 4.86 \\
Val. & d9s6 & 770 & 0.13 & 10.91 & 88.96 & 6.4 & 37.9 & 1.79 / 2.31 / 3.48 \\
\bottomrule
\end{tabular}

\vspace{2pt}{\footnotesize Border: the person touches the image border. Legs hidden: the person does not touch the border and a visible ankle lies more than 5\% of the box height below the box bottom. $\rho$ is the range-axis center of the radar box, a proxy for the person's range rather than the Euclidean slant range. No test window contains more than three people, and only seven contain three; two validation windows of d6s4 contain four.}
\end{table*}

\emph{Training.} \rev{We train from scratch with AdamW~\cite{adamw} (learning rate $2\times10^{-4}$, weight decay 0.05) at batch size 32 for 15 epochs, with a one-epoch warm-up, cosine decay, global gradient clipping at 1.0, and an exponential moving average of the weights with decay 0.9998, which alone is evaluated.} Training uses bfloat16 mixed precision, weight decay only on convolution, linear and attention projection weights, and cosine decay to $10^{-6}$ over 15 epochs of 1,233 updates each. The augmentation shifts the log heatmaps globally by up to $\pm0.3\ln10$ per sample and adds Gaussian noise of standard deviation 0.02. Each denoising position is perturbed by Gaussian noise with 1.5 times the 68th-percentile localization error of the model in range and in the two tangents, its log-size by Gaussian noise of standard deviation 0.3, and its keypoint fractions by uniform noise of $\pm0.1$. A small fraction of the denoising queries are placed uniformly in the field of view but more than the response radius from every person laterally and more than 0.3\,m from every person in range, and are trained as background. Epochs are selected on validation from the moving-average weights\rev{: the selected epoch has the highest validation pose AP among those whose raw observed-box AP is within 0.3 of the best,} with ties broken toward the earlier epoch; the selected epochs are 15, 14 and 13 for seeds 1--3. All three seeds were trained on one server with PyTorch 2.8.

\section{Confident Boxes and Duplicates}
\label{S-app:counts}
Table~\ref{S-tab:counts} gives the counts behind Section~\ref{sec:duplicates} of the main paper. A box is confident when its score exceeds 0.5, a threshold fixed before the counts were made. The HOMER rows are the three seeds of Table~\ref{tab:main}; the other rows of our models are the ablations of Table~\ref{tab:ablation}. Counts at one threshold describe the outputs; they do not decompose AP, which integrates over all thresholds.

\begin{table}[!tbp]\centering
\caption{Confident boxes (score above 0.5) on the test split.}\label{S-tab:counts}
\footnotesize\setlength{\tabcolsep}{1.8pt}
\begin{tabular}{@{}lcccccc@{}}
\toprule
 & Boxes per & Exact & Multi- & Dupl.\ per & & \\
Model & person & count (\%) & box (\%) & 1,000 & TP & FP\\
\midrule
\multicolumn{7}{@{}l}{\emph{Unseen room (d8)}}\\
HOMER (3 seeds) & 1.45 & 47.5 & 54.8 & 74 & 52{,}164 & 47{,}708\\
\quad 12 frames & 1.37 & 53.4 & 47.5 & 90 & 51{,}940 & 42{,}403\\
REXO (3 runs) & 1.09 & 64.1 & 33.5 & $<$1 & 46{,}638 & 28{,}655\\
Ablation, seed 1 & 1.39 & 52.7 & 47.2 & 70 & 49{,}918 & 46{,}262\\
\quad no suppression & 1.77 & 30.4 & 65.8 & 372 & 48{,}862 & 73{,}574\\
\midrule
\multicolumn{7}{@{}l}{\emph{Seen sessions}}\\
HOMER (3 seeds) & 1.28 & 70.1 & 27.7 & 76 & 26{,}689 & 16{,}434\\
\quad 12 frames & 1.22 & 72.6 & 23.6 & 70 & 26{,}189 & 14{,}690\\
REXO (3 runs) & 0.98 & 94.1 & 1.8 & $<$1 & 28{,}155 & 4{,}642\\
Ablation, seed 1 & 1.17 & 75.4 & 19.9 & 62 & 25{,}006 & 14{,}178\\
\quad no suppression & 1.72 & 46.9 & 51.6 & 361 & 24{,}534 & 33{,}062\\
\bottomrule
\end{tabular}

\vspace{2pt}\parbox{\columnwidth}{\footnotesize Boxes per person: confident boxes divided by labeled people. Exact count: windows whose number of confident boxes equals the number of people. Multi-box: single-person windows with two or more confident boxes. Duplicates: confident boxes beyond the first that overlap the same labeled person at IoU above 0.5, per 1,000 labeled people. TP and FP: greedy matching in score order at IoU $\ge0.5$ on the clipped observed box. Means over training seeds or inference runs. Ablation, seed 1: the 12-frame configuration with radar-space supervision of the lower block of Table~\ref{tab:ablation}, with and without peak suppression.}
\end{table}

\section{Comparison with REXO by Stratum}
\label{S-app:strata}
\rev{Fig.~\ref{fig:strata}} \rev{of the main paper} splits the box comparison of Table~\ref{tab:main} by part and number of people, \rev{by range and truncation,} and by session. \rev{AP is the clipped observed-box AP, averaged over the three seeds of HOMER and the three inferences of REXO; strata that were scored only for the 12-frame variant of Table~\ref{tab:ablation} use its three seeds. The range $\rho$ of a window is the median range of the radar boxes of its labeled people, and the two windows beyond 6\,m are omitted; a window touches the border when at least one labeled person does. In the eight multi-person windows of the seen sessions, 105 of the 2,000 replicates contain no labeled person and are dropped.} The intervals resample contiguous segments within each stratum, as in the main paper. Because the test sessions of the seen rooms contain almost no multi-person windows, the table also gives the validation sessions of those rooms; they were used to choose the epoch and are not an independent test.

\section{Zero-Shot Transfer to the P1 Room}
\label{S-app:p1}
Part P1 of MMVR was recorded with a single subject in an open-foreground room that does not occur in P2~\cite{mmvr}. We applied the three checkpoints of the 12-frame variant (Table~\ref{tab:ablation}), REXO with its three inference seeds and RETR, all trained on P2S2, to the two official P1 test subsets without training or selecting anything on P1. The floor constants are those used for d8, the medians over the four training rooms (Sec.~\ref{S-app:floor}), which assumes a similar mounting height of the radar. Table~\ref{S-tab:p1} gives the results. Its mean is above REXO on both subsets, by 2.16 and 6.23 AP; all three seeds are above every REXO inference on the second subset, whereas one seed, at 18.74, is below REXO on the first. We report no interval, and the two subsets overlap.

\begin{table}[!tbp]\centering
\caption{Zero-shot transfer to the P1 room.}\label{S-tab:p1}
\footnotesize\setlength{\tabcolsep}{1.9pt}
\begin{tabular}{@{}lcccccc@{}}
\toprule
 & & \multicolumn{4}{c}{Box AP} & Pose AP\\
\cmidrule(lr){3-6}\cmidrule(l){7-7}
Subset & Windows & RETR & REXO & Ours & Seeds 1/2/3 & Ours\\
\midrule
P1S1 & 10{,}785 & 13.12 & 19.05 & 21.21$\pm$2.22 & 23.03/18.74/21.88 & 23.08$\pm$3.06\\
P1S2 & 13{,}238 & 13.26 & 19.51 & 25.74$\pm$2.64 & 28.57/23.34/25.32 & 17.93$\pm$1.73\\
\bottomrule
\end{tabular}

\vspace{2pt}\parbox{\columnwidth}{\footnotesize Ours: the 12-frame variant of Table~\ref{tab:ablation}. All models are P2S2 checkpoints, applied without training or selection on P1. The two official P1 test subsets overlap and are not independent. Box AP: clipped observed box; pose AP: MMVR Eq.~(3) with box area. REXO: mean of three inferences.}
\end{table}

\section{Typical Outcomes}
\label{S-app:outcomes}
\begin{figure}[!tbp]\centering
\includegraphics[width=\columnwidth]{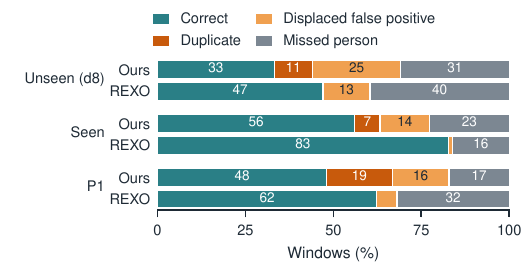}
\caption{\rev{Share of windows in each outcome class: 52,115 windows (d8), 33,562 (seen) and 22,223 (P1, the union of the two test subsets). Ours: seed 1 of the 12-frame variant of Table~\ref{tab:ablation}; REXO: first inference. Confident boxes and matching as in Table~\ref{S-tab:counts}. Correct: every person matched and no false positive. Duplicate: no person missed and at least one false positive overlapping a matched person at IoU above 0.5. Displaced false positive: no person missed and every false positive at IoU 0.5 or less with every person. Missed person: at least one person unmatched.}}\label{S-fig:outcomes}\end{figure}

To see what the counts of Table~\ref{S-tab:counts} look like in the data, we classified every window by the outcome of its confident boxes (\rev{Fig.~\ref{S-fig:outcomes}}), for seed 1 of the 12-frame variant and the first REXO inference and with the matching of Table~\ref{S-tab:counts}; the totals of matched people and errors equal those behind that table. The two models fail in different ways. In the unseen room ours ends with an extra box, on the person or beside it, in 36\% of the windows against 13\% for REXO, which almost never places two boxes on one person, while REXO misses someone in 40\% of the windows against 31\% for ours. Fewer misses and more extra boxes is the pattern behind the higher recall and the lower count accuracy of HOMER in Table~\ref{S-tab:counts}, and REXO ends exactly right in more windows in every part of the test.

\begin{figure*}[!tbp]\centering
\includegraphics[width=\textwidth]{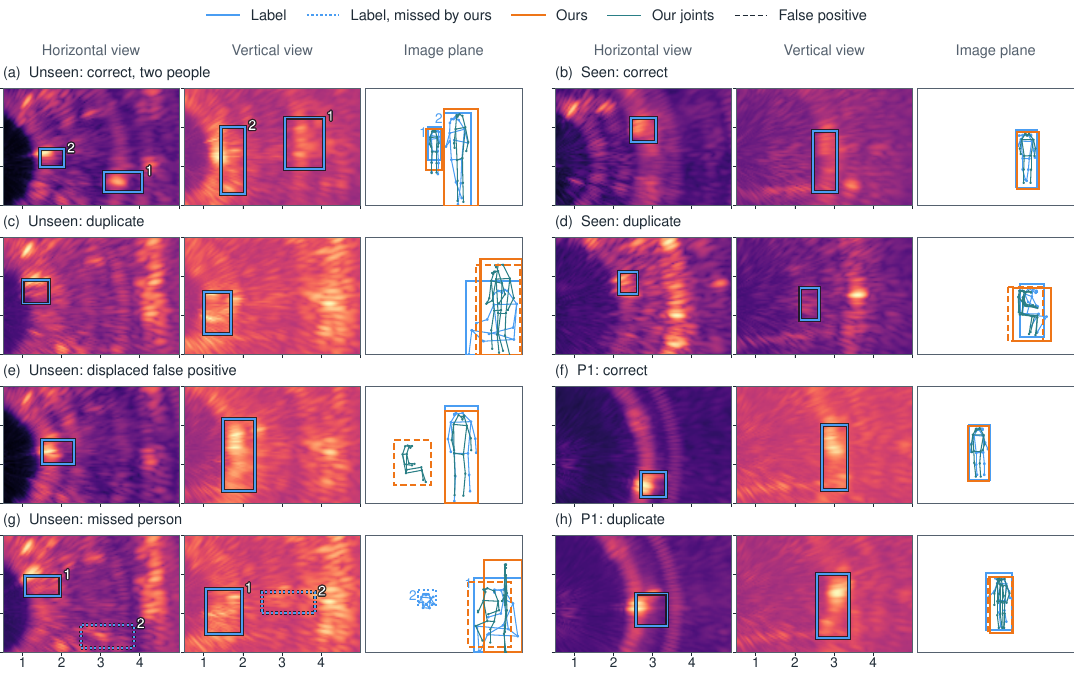}
\caption{\rev{Typical outcomes of ours (seed 1 of the 12-frame variant), one window per class and part. For each window: horizontal and vertical heatmap of the current frame in dB below its maximum (25\,dB shown; range 0.5--5\,m to the right, lateral position and height above the radar upward), with the label boxes in radar space, and the image plane with the labels and our confident boxes and joints; false positives are dashed.}}\label{S-fig:gallery}\end{figure*}

\rev{Fig.~\ref{S-fig:gallery} shows one window of each class of ours in each part, the window at the middle of the class in a fixed order, chosen before any image was viewed.}

\section{Peak Suppression in the Locked Windows}
\label{S-app:traces}
\begin{figure*}[!tbp]\centering
\includegraphics[width=\textwidth]{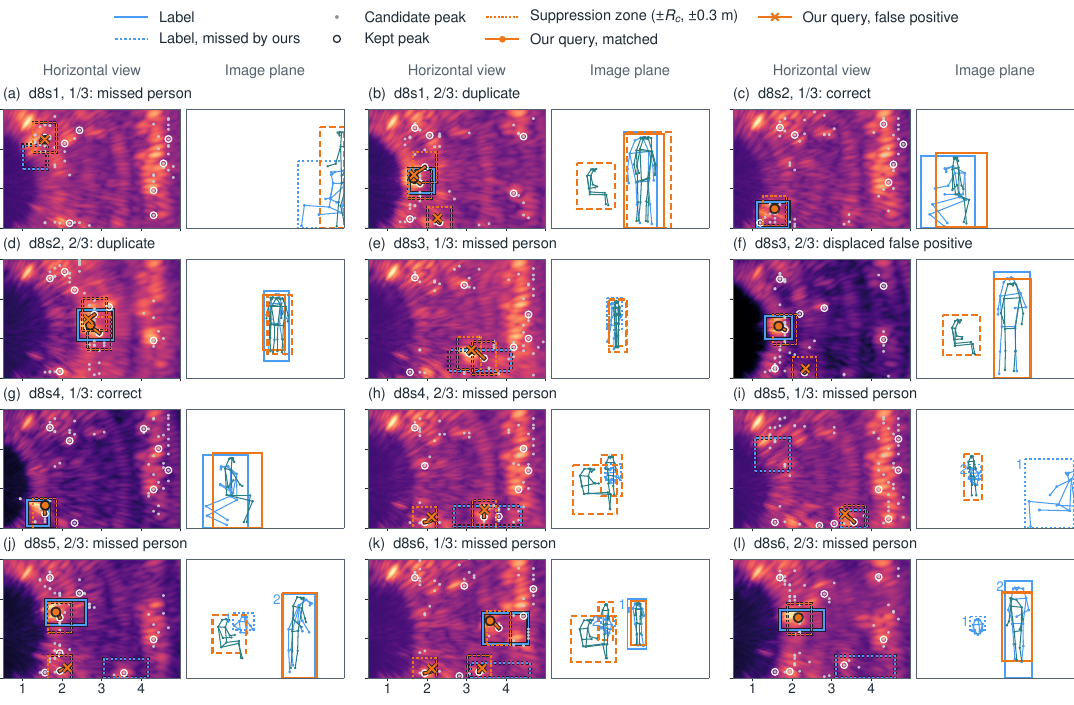}
\caption{\rev{Peak suppression on the twelve windows of the unseen room fixed before testing, first seed of the 12-frame variant, drawn as in Fig.~\ref{fig:suppression} of the main paper. Horizontal heatmaps in dB below the frame maximum (25\,dB shown); labels missed by ours are dotted.}}\label{S-fig:traces}\end{figure*}

Fig.~\ref{fig:suppression} of the main paper draws three of the twelve windows of the unseen room that were fixed before any result was seen, the windows at one third and two thirds of each d8 session, for the first seed of the 12-frame variant; the outputs are those of the frozen predictions. In every one of the twelve windows eight peaks survive, so no query is filled from discarded candidates, and every confident query starts on a kept peak. \rev{Fig.~\ref{S-fig:traces} draws all twelve windows.}

\begin{figure*}[!tbp]\centering
\includegraphics[width=\textwidth]{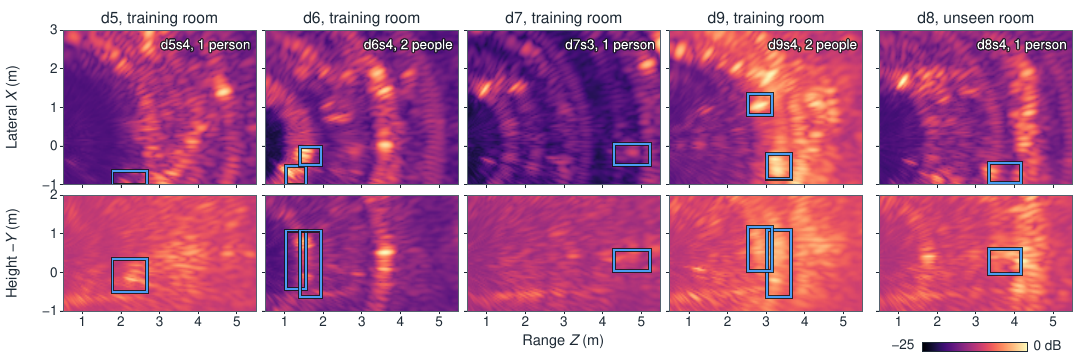}
\caption{\rev{One window per room, the middle frame of the middle segment of the middle session, fixed by rule before plotting, in dB below the maximum of each map: horizontal heatmap (lateral position against range, top) and vertical heatmap (height above the radar against range, bottom), with the label boxes. The room medians over 10,000--12,000 frames, 2,000 sampled from each session (all 430 of d6s5), are in Fig.~\ref{fig:rooms} of the main paper.}}\label{S-fig:rooms}\end{figure*}

The twelve windows hold 16 labeled people and 21 confident queries, of which 8 are matched and 13 are false positives, of three kinds. Four are second queries on one person: in two windows two queries start 0.41 and 0.47\,m apart laterally, just beyond $R_c$ (0.38 and 0.41\,m), and in a third 0.38\,m apart in range, just beyond the 0.3\,m gate; in all three the decoder brings them within 0.15\,m of each other. The suppression separates the peaks as designed; the regression of the decoder, which the search windows do not bound (Sec.~\ref{S-app:rc}), merges them. Five sit at one fixed place, 0.72--0.76\,m to the left at 2.1--2.3\,m range, in five windows from five sessions and whoever else is in the room. Nobody is labeled there, the heatmap there is among the strongest 5\% of the map in 11 of the 12 windows, a peak is kept there in 9 of them, and HOMER draws a seated figure at it. A fixed reflector of this kind belongs to the room, so a model trained in other rooms has not learned to ignore it. \rev{That this one belongs to d8 can be seen without any model (Fig.~\ref{fig:rooms} of the main paper, which shows the horizontal and vertical medians of every room). The per-pixel median of the heatmaps over 2,000 frames of each session, taken without reading any label, is high at $X=-0.82$\,m, $Z=2.05$\,m in all six d8 sessions, among the strongest 1\% of the field drawn in Fig.~\ref{fig:suppression} and 7.7--9.1\,dB above its median, whereas the same place lies between the 20th and 55th percentiles in the four training rooms. Fig.~\ref{S-fig:rooms} shows one window of each room.} The remaining four lie near a labeled person, within 0.45\,m laterally and in range, with an image box that misses the label at IoU 0.5.

\end{document}